\documentclass[11pt]{article}{}
\pdfoutput=1

\newcommand{\doctitle}{Pattern formation in dense populations studied by inference of nonlinear diffusion-reaction mechanisms}
\newcommand{\docauthor}{S. Srivastava et.al.}
\newcommand{\docaffil}{Applied Physics, Mathematics, Mechanical Engineering, and Michigan Institute for Computational Discovery \& Engineering\\ University of Michigan}

\newcommand{\docheader}{}
\newcommand{\docfooter}{University of Michigan}

\newcommand{\pdftitle}{reaction-diffusion}
\usepackage{main}

\usepackage{xpatch}
\xapptocmd\appendices{%
  \crefalias{section}{appendix}%
}{}{\PatchFailed}
\crefname{section}{Sec.}{Secs.}
\Crefname{section}{Sec.}{Secs.}
\crefname{equation}{Eq.}{Eqs.}
\Crefname{equation}{Eq.}{Eqs.}
\creflabelformat{equation}{#2(#1)#3}
\crefname{figure}{Fig.}{Figs.}
\Crefname{figure}{Fig.}{Figs.}
\creflabelformat{figure}{#2#1#3}
\crefname{table}{Tab.}{Tabs.}
\Crefname{table}{Tab.}{Tabs.}
\creflabelformat{table}{#2#1#3}
\Crefname{appendix}{App.}{Apps.}
\crefname{appendix}{App.}{Apps.}

\begin{document}
\pretitle{\begin{center}\vskip -80pt}%
\title{\Large\doctitle}
\posttitle{\end{center}}
\preauthor{\begin{center} \vskip -2pt}
\author[1,3]{Siddhartha Srivastava}
\author[1,2,3]{Krishna Garikipati\thanks{Corresponding author at: Department of Mechanical Engineering, University of Michigan, United States. \emph{E-mail address}: krishna@umich.edu (K. Garikipati).}} 
\affil[1]{Department of Mechanical Engineering, University of Michigan, United States}
\affil[2]{Department of Mathematics, University of Michigan, United States}
\affil[3]{Michigan Institute for Computational Discovery \& Engineering, University of Michigan, United States}
\postauthor{\end{center} \vskip -20pt}
\predate{\begin{center} \vskip -0pt}
\date{} 
\postdate{\end{center} \vskip -40pt}%
\maketitle
\begin{abstract}
\pagestyle{abstract}

Reaction-diffusion systems have been proposed as a model for pattern formation and morphogenesis. The Fickian diffusion typically employed in these constructions model the Brownian motion of particles.  The biological and chemical elements that form the basis of this process, like cells and proteins,  occupy finite mass and volume and interact during migration.  We propose a Reaction-diffusion system with Maxwell-Stefan formulation to construct the diffusive flux. This formulation relies on inter-species force balance and provides a more realistic model for interacting elements. We also present a variational system inference-based technique to extract these models from spatiotemporal data for these processes. We show that the inferred models can capture the characteristics of local Turing instability that instigates the pattern formation process. Moreover, the equilibrium solutions of the inferred models form similar patterns to the observed data. 
\end{abstract}

\section{Introduction}

Biological patterns and morphogenesis are emergent phenomena with discernibly coherent structures forming at larger length scales, spawning out of local mechanochemical interactions of natural materials. A large class of PDEs have been proposed to explain this phenomenon. These models have been reviewed in a previous communication \cite{garikipati2017perspectives}. Among these models, the Reaction-Diffusion (RD) model is one of the most widely studied in literature \cite{konda2010}. The seminal work of Alan Turing\cite{turing1990chemical} laid out the mathematical understanding of how specific spatial modes are destabilized in an RD system, leading to a heterogeneous equilibrium solution. The physical understanding of this mechanism regards the morphogens as short-range activators coupled with long-range inhibitors. A local positive perturbation in the activator leads to a localized growth in activators and inhibitors. Meanwhile, the faster diffusion of the inhibitors leads to a decay in activators and inhibitors in the neighboring locations. In this work, we take a mathematical outlook on the RD system, albeit guided by our understanding of the physical system. In particular, we focus on the diffusion aspect of this model. 

A lot of development in the study of RD systems, starting from Turing's original work\cite{turing1990chemical}, has been with Fickian diffusion with varying self-diffusivities of different components \cite{konda2010, green2015positional}. 
Some studies \cite{madzvamuse2015cross, vanag2009cross} have also considered cross-diffusion as a generalization to the diagonal dominant diffusion. Fickian diffusion\cite{Gorban2010QuasichemicalMO} posits that the flux of a chemical component is directly proportional to the negative gradient of its concentration. At the microscopic scale, this transport models the \textit{Brownian motion} that is the stochastic collision-driven motion of particles. The cumulative migratory behavior of these particles translates to a concentration gradient equalizing flow of the density.  This linear relationship of flux and concentration gradient works reasonably well for approximating diffusion processes in binary mixtures and under dilute limits \cite{KRISHNA1997861}. On the other hand, extending this model to more complicated scenarios entails including correction terms in the diffusivity coefficient \cite{darken1948diffusion,vignes1966diffusion,DAGOSTINO2011}.
Often, it is not even possible to describe the behavior with these corrections while also maintaining thermodynamic admissibility. Duncan and Toor\cite{duncan1962experimental} designed an experiment with a ternary mixture of gases that showcased behavior like osmotic diffusion, reverse diffusion and diffusion barrier where the apparent driving force seems to be along the direction of gradient; such phenomena are not captured by Fickian diffusion. The Maxwell-Stefan (MS) approach to multicomponent diffusion is based on the kinetic theory and molecular force balances \cite{Bothe2011}. It has been applied to various physical systems such as intracellular organization of biomolecules\cite{ramm2021diffusiophoretic}, protein adsorption\cite{yang2016dynamic, sun2008analysis}, flow of solutes across polymer membranes\cite{membranes12100942} and dispersion of atmospheric aerosols\cite{fowler2018maxwell}. This mechanism can better describe the interaction of elements in specific biological systems, for instance, cellular migration in tissue formation.

In this work, we have developed an MS diffusion-based RD system. A similar model has been developed by Herberg et.al.\cite{herberg2017reaction} where they have considered a mass-conserving transport with reactive flow; however, they have not considered pattern-forming reactive dynamics. We also provide a mathematical framework to obtain these models in a data-driven fashion. We believe this approach will also help alleviate the problem in the experimental measurements of MS diffusivity coefficients\cite{giacomazzi2008review}.  Our approach leverages Variational System Identification (VSI) techniques that adopt a regression-based methodology. This methodology involves evaluating a set of potential operators within partial differential equations (PDEs) using available data. Subsequently, the parameter estimation problem is reformulated as a nonlinear regression problem. This transformation can be achieved through the construction of operators, either in a strong form of PDE, such as finite difference representations as seen in the Sparse Identification of Nonlinear Dynamics  (SINDy) approach\cite{sindy}, or in a weak or variational form of the PDE as is done in VSI methods. Both of these operator representations impose rigorous demands on the quality and quantity of data necessary for accurate operator construction. However, using weak forms offers distinct advantages, including the appearance of boundary conditions within the formulation and reduced prerequisites concerning spatial derivatives. PDE-constrained optimization techniques have also been employed for parameter identification\cite{wang2021variational}. These techniques, unlike VSI, minimize the difference 
in the PDE's predicted solution and the data field. Therefore, they enjoy the advantage of being much more robust against noisy data, but at a computational cost. However, in the study of pattern formation, the equilibrium solution has a strong sensitivity to initial conditions, making PDE-constrained optimization a poor choice for such problems. In the case of PDE-constrained optimization, this problem is further exacerbated when the forward solution of models does not converge due to poor guesses for parameters. The mathematically simplified nature of the regression approach, particularly when working with weak forms, renders it amenable to a wide array of operator selection techniques. These include regularization, stepwise regression, and genetic algorithms, facilitating the identification of a parsimonious model (PDE with minimal parameters) from an extensive set of potential operators\cite{wang2021variational}. 
Similar techniques have been previously studied across diverse domains, including constitutive theories\cite{wang2021variational, nikolov2022ogden},epidemiology\cite{wang_system_2021}, and pattern formation\cite{wang_variational_2019}. 

We have made the following contributions in this work: 
\begin{itemize}
    \item We have proposed a new model for studying pattern formation using an MS-based RD system. 
    \item We show that MS-based RD systems can capture a rich set of dynamics that generalizes the ones produced by Fick-based RD systems.
    \item We have formulated a VSI framework for extracting the RD model from spatiotemporal data. 
    \item Our approach captures the salient features of the Turing instability, and inferred models can produce similar patterns to the original data.
    \item We show that temporally refined data near the onset of instability results in better recovery of stability diagrams. Meanwhile, including longer time ranges in the data results in better recovery of patterns
\end{itemize}

In the next section, we will introduce the theory for the transport of multicomponent systems and Maxwell-Stefan diffusion. In \cref{sec:vsi}, we develop the VSI-based computational framework for inferring MS-based RD systems from spatiotemporal data of concentration fields. In \cref{sec:pattern}, we show pattern formation with MS-based RD systems and employ our data-driven inference approach for identifying these systems. Finally, we present the concluding remarks in \cref{sec:conclusion}.

\section{Transport of multicomponent system}\label{sec:transport_review}

Let $\Omega \subset \mathbb{R}^m$ be an open bounded domain with boundary $\partial \Omega$ with outer normal field $\boldsymbol{n}$. We consider a mixture with $n\geq 2$ components. Each constituent has a molar mass $M_k > 0$, individual mass density $\rho_k \geq 0$ and a velocity $\boldsymbol{u}_i$ satisfying $\boldsymbol{u}_i\cdot \boldsymbol{n} = 0$ on the boundary $\partial \Omega$. The mass balance of each component is given by: 
\begin{align}\label{eq_mass_comp_transport}
    \frac{\partial \rho_i}{\partial t} + \nabla\cdot \rho_i \boldsymbol{u}_i = M_i r_i \quad \text{ in } \Omega \text{ for } t\geq 0
\end{align}

where $r_i$ is the rate of production of $i^{\text{th}}$ species. Note that $r_i $ is \textit{positivity-preserving} i.e. if $\rho_i = 0$ then $r_i\geq 0$. The total mass balance is obtained by summing over the balance equation of each component: 
\begin{align}\label{eq_mass_transport}
    \frac{\partial \rho}{\partial t} + \nabla\cdot \rho \boldsymbol{u} =\sum_i M_i r_i, \quad \text{ with } \boldsymbol{u} \cdot \boldsymbol{n} = 0 
\end{align}
where $\rho = \sum_i \rho_i$ is the total mass. The barycentric velocity, $\boldsymbol{u}$, is estimated as  $\rho \boldsymbol{u}= \sum_i \rho_i \boldsymbol{u}_i$. Transport is also studied in terms of other quantities like mass faction $y_i = \rho_i/\rho$,  molar concentration, $c_i=\rho_i/M_i$, and molar factions $x_i=c_i/c_{\text{tot}}$ where the total concentration is estimated as $c_{\text{tot}} = \sum_i c_i$. The transport of these quantities can be obtained from the mass transport equation. For instance, combining \eqref{eq_mass_comp_transport}, \eqref{eq_mass_transport} and the definition of mass faction gives: 
\begin{align}\label{eq_mass_faction_transport}
\rho (\partial_t y_i +   \boldsymbol{u} \cdot\nabla y_i )  + \nabla\cdot \overline{\boldsymbol{J}}_i  = M_i r_i - y_i\sum_iM_i r_i, \quad \text{ with } \overline{\boldsymbol{J}}_i \cdot \boldsymbol{n} = 0
\end{align}
where $\overline{\boldsymbol{J}}_i = \rho_i(u_i - u)$. 
The transport of molar concentration is obtained as follows: 
\begin{align}\label{eq:ci_transport}
\partial_t c_i +\nabla\cdot c_i \boldsymbol{u}_i =  r_i
\end{align}
The transport of total concentration is obtained by adding the transport of individual concentrations: 
\begin{align}\label{eq:ctot_transport}
    \partial_t c_{\text{tot}} +\nabla\cdot  c_{\text{tot}} \boldsymbol{v} = \sum_i r_i, \quad \text{ with } \boldsymbol{v} \cdot \boldsymbol{n} = 0
\end{align}
where molar averaged velocity, $\boldsymbol{v}$ is evaluated as $c_{\text{tot}}\boldsymbol{u} = \sum_i c_i \boldsymbol{u}$. 
Finally the transport of molar factions are obtained from \eqref{eq:ci_transport} and \eqref{eq:ctot_transport}:
\begin{align}\label{eq:xi_transport}
c_{\text{tot}}(\partial_t x_i  +   \boldsymbol{v} \cdot\nabla x_i) +   \nabla\cdot \boldsymbol{J}   =  r_i - x_i\sum_i r_i, \quad \text{ with } \boldsymbol{J}_i \cdot \boldsymbol{n} = 0
\end{align}
where $\boldsymbol{J}_i =   c_i (\boldsymbol{u}_i-\boldsymbol{v})$. 
It should be remarked that previous studies on non-reactive systems\cite{KRISHNA1997861} assume $c_{\text{tot}}$ to be constant over time. It can be observed from \eqref{eq:ctot_transport} that this requires: (1) $\sum_i r_i = 0$, i.e. there is no net production of material in the system, (2)  $\nabla\cdot  c_{\text{tot}} \boldsymbol{v}=0$ that can be achieved by fixing a frame of reference moving with molar averaged velocity. 
meanwhile, for reactive flows \cite{herberg2017reaction}, it is assumed that $\rho$ is constant over time, requiring (1) $\sum_i M_i r_i = 0$ i.e. there is no net production of mass in the system, (2) $\nabla\cdot \rho \boldsymbol{u}=0$ that can be achieved by fixing a frame of reference moving with barycentric velocity. 

For this work, we will study the transport of molar factions and fix $c_{\text{tot}} = 1$. It should be noted that in this work, molar factions and molar concentration will have the same numerical value in the units of $c_{\text{tot}}$. As mentioned earlier, we will impose a constant $c_{\text{tot}}$ by setting $\sum_i r_i = 0$ that we refer to as the \textit{no-net production} property. Moreover, we fix the frame of reference with respect to molar averaged velocity, and $\boldsymbol{v}=0 \implies \sum_i J_i = 0$. From here on, we will also refer to flux as $J_i =  c_i \boldsymbol{u}_i$.   With these remarks, the system under consideration is rewritten as follows: 
\begin{align}\label{eq:transport}
\partial_t c_i +\nabla\cdot  \boldsymbol{J}_i =  r_i, \quad \text{ such that } \sum_i r_i=0
\end{align}

The first prominent equation of diffusion is Fick’s law. This law states that for a single-component system, the diffusion flux is proportional to the negative gradient of the concentration.
\begin{equation}
    \boldsymbol{J} = - D \nabla c
\end{equation}
Therefore, we get that the time derivative of the concentration is the negative of the divergence of the flux:
\begin{equation}
    \frac{\partial c}{\partial t} = - \nabla \cdot \boldsymbol{J} = \nabla \cdot D\nabla c
\end{equation}



This theory can be phenomenologically extended to the multicomponent case by assuming flux to be given as a linear function of the gradient fields as follows:  
\begin{equation}
    \boldsymbol{J}_i  = - \sum_{j=1}^{n}  D_{ij} \nabla c_j
\end{equation}
where the diffusivity tensor, $\bD\equiv \bD([c_1,\cdots c_n])$ consists of non-zero off-diagonal terms:
\begin{equation}
    \bD = \begin{bmatrix}
D_{11} & D_{12} & \cdots & D_{1n}\\ 
D_{21} & D_{22} & \cdots & D_{2n}\\
\vdots & \vdots & \ddots &\vdots \\
D_{n1} & D_{n2} & \cdots & D_{nn}
\end{bmatrix}
\end{equation}
This results in the following RD model:
\begin{align}\label{eq:reaction_diffusion}
    \partial_t c_i = \sum_j \nabla\cdot D_{ij} \nabla c_j +  r_i
\end{align}
However, it has been proven that this phenomenology for diffusive fluxes is thermodynamically inconsistent\cite{BOTHE2023103818}. In particular, (1)
mass conservation is only satisfied if all diffusivities are the same. (2) The second law is satisfied only in very special cases. In the next section, we present Maxwell-Stefan diffusion that alleviates these deficiencies.

\subsection{Maxwell-Stefan diffusion}

The MS equations rely on inter-species force balances. More precisely, it is assumed that the thermodynamical driving force $d_i$ is in local equilibrium with the total friction force. The friction force on 1 mole of particle $i$ due to particle $j$ is modeled as
\begin{align}
    f_{ij} = \frac{1}{\mathcal{D}_{ij}}c_j(u_i - u_j)
\end{align}
where the $\mathcal{D}_{ij}>0$ has a physical significance of an inverse drag coefficient and $\mathcal{D}_{ij} = \mathcal{D}_{ji}$ is a natural condition due to force balance. We remark that the formulations presented in the literature show some variations in the model for net frictional forces, for instance, Krishna and Wesselingh\cite{KRISHNA1997861} consider $f_{i,j} \propto x_j(u_i-u_j)$, meanwhile Herberg et.al. \cite{herberg2017reaction} consider $f_{i,j} \propto y_j(u_i-u_j)$. 

The net force balance on particles results in the following:
\begin{equation}
    d_i = - \sum_{j\neq i} \frac{c_i c_j}{\mathcal{D}_{ij}} (u_i - u_j) = - \sum_{j\neq i} \frac{c_jJ_i - c_iJ_j}{\mathcal{D}_{ij}}
\end{equation}

Note that this relation satisfies $\sum_i d_i = 0$. There have been various models for driving forces proposed for specific systems\cite{bothe2015continuum}. In this work we will assume that $d_i = c_i\nabla \mu_i$, with the standard chemical potentials $\mu_i = \log(\gamma_i c_i)$ for the potentials. Here $\gamma_i > 0$ are the so-called activity coefficients. This results in $d_i = \nabla c_i$. These relations are then used to estimate $J_i$ in terms of the thermodynamic forces.

The special case of equimolar transport is well-studied in literature\cite{KRISHNA1997861}. Mathematically, it implies that $\sum_i J_i = 0$ and $c_\text{tot}$ is constant. Consequently, $ \sum_i \nabla c_i = \nabla c_\text{tot} = 0$. This allows us to rewrite the system with $n-1$-independent variables, ($c_1, \cdots c_{n-1}$): 
\begin{align}\label{eq:MS2flux_reduced}
    \bB \begin{bmatrix}
    J_1 \\
    \vdots\\
    J_{n-1}
    \end{bmatrix} =  - \begin{bmatrix}
    \nabla c_1 \\
    \vdots\\
    \nabla c_{n-1}
    \end{bmatrix}
\end{align}
where 
\begin{align*}
    {B}_{ij} = \left\{\begin{matrix}
-c_i\left(  \frac{1}{\mathcal{D}_{ij}} - \frac{1}{\mathcal{D}_{1n}}\right)& i\neq j\\ 
\frac{1}{\mathcal{D}_{in}} + \sum_{k\neq i} c_k\left(  \frac{1}{\mathcal{D}_{ik}} - \frac{1}{\mathcal{D}_{in}}\right) & i=j
\end{matrix}\right.
\end{align*}
where we have used the facts: (1)$\sum_{1}^n J_i = 0 $, (2)$\sum_{1}^n c_i = c_\text{tot} = 1 $, and (3) $\sum_{1}^n \nabla c_i = 0$ 

\textit{Example}: Consider a non-reactive ternary system. In this case, we have two independent equations where the $\bB$ matrix is given as: 
\begin{equation*}
    \bB = \begin{bmatrix}
    \frac{1}{\mathcal{D}_{13}} + c_2 \left( \frac{1}{\mathcal{D}_{12}} - \frac{1}{\mathcal{D}_{13}} \right) & - c_1 \left( \frac{1}{\mathcal{D}_{12}} - \frac{1}{\mathcal{D}_{13}} \right)\\
    - c_2 \left( \frac{1}{\mathcal{D}_{12}} - \frac{1}{\mathcal{D}_{23}} \right) &  \frac{1}{\mathcal{D}_{23}} + c_1 \left( \frac{1}{\mathcal{D}_{12}} - \frac{1}{\mathcal{D}_{23}} \right) 
    \end{bmatrix}
\end{equation*}
Noting that $\det \bB = (\mathcal{D}_{13}\mathcal{D}_{23})^{-1} + c_2 \mathcal{D}_{23}^{-1} \left( \mathcal{D}_{12}^{-1} - \mathcal{D}_{13}^{-1} \right) + c_1 \mathcal{D}_{13}^{-1} \left( \mathcal{D}_{12}^{-1} - \mathcal{D}_{23}^{-1} \right)$
\begin{equation}\label{eq:MS2diffusion}
    \bB^{-1} = \frac{1}{\det \bB}\begin{bmatrix}
    \frac{1}{\mathcal{D}_{23}} + c_1 \left( \frac{1}{\mathcal{D}_{12}} - \frac{1}{\mathcal{D}_{23}} \right) &  c_1 \left( \frac{1}{\mathcal{D}_{12}} - \frac{1}{\mathcal{D}_{13}} \right)\\
     c_2 \left( \frac{1}{\mathcal{D}_{12}} - \frac{1}{\mathcal{D}_{23}} \right) &  \frac{1}{\mathcal{D}_{13}} + c_2 \left( \frac{1}{\mathcal{D}_{12}} - \frac{1}{\mathcal{D}_{13}} \right) 
    \end{bmatrix}
\end{equation}
The evolution equation for this reduced system is given as: 
\begin{align}\label{eq:MS2_reduced}
\frac{\partial c_i}{\partial t} 
 = \sum_{j=1}^{2}\nabla \cdot   
[\bB^{-1}]_{ij} \nabla c_j  
\end{align}
Given the similarity between the forms of MS and Fickian diffusivities from here on, we will denote with ${D}_{ij} = [\bB^{-1}]_{ij}$, noting that these coefficients are now considered to be functions of the molar concentration. 
\begin{figure}
    \centering
    \includegraphics[width=0.85\linewidth]{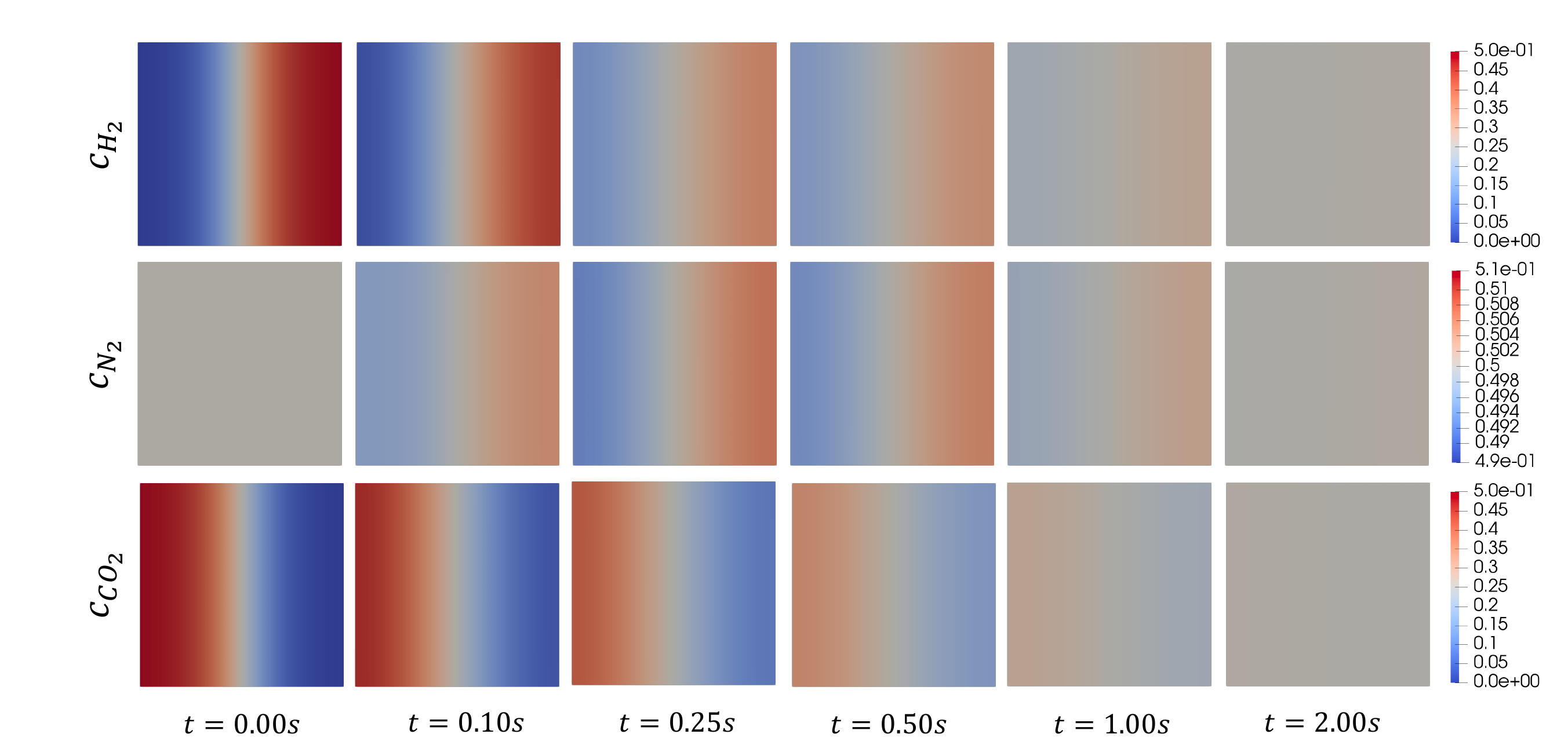}
    \caption{
    Evolution of molar concentrations $c_{H_2}$ and $c_{N_2}$ over $t\in [0,2]$.}
    \label{fig:ms_duncan}
\end{figure}

\begin{figure}

    \centering
    \includegraphics[width=0.85\linewidth]{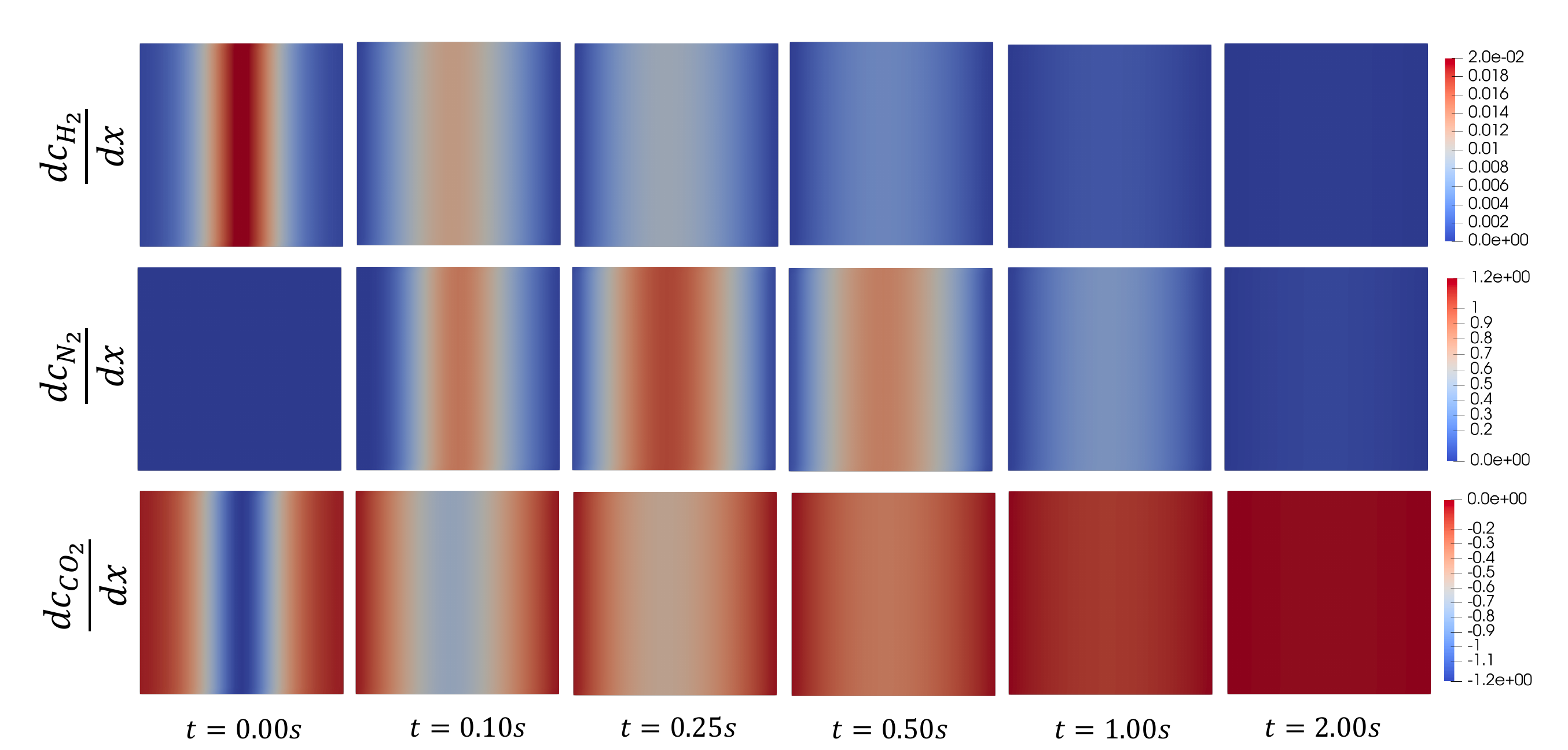}
    \caption{
    Evolution of gradients of molar concentrations $c_{H_2}$ and $c_{N_2}$ over $t\in [0,2]$.
    }
    \label{fig:ms_duncan_grad}
\end{figure}

\subsection{Case study for non-reactive equimolar transport}\label{sec:dunan_n_toor}
Duncan and Toor's experiment\cite{duncan1962experimental} provides a canonical illustration for the generalizability of MS formulation over Fickian fluxes. The experimental set-up consisted of two-bulb diffusion cells with an ideal ternary gas mixture of (1) $H_2$ (2 ) $N_2$ (3) $CO_2$. , where bulb 1 consisted of equal moles of $H_2$ and $N_2$ mixture and bulb 2 consisted of equal moles of $N_2$ and $CO_2$ mixture.  The mixtures in the bulbs were allowed to diffuse into each other via a capillary. The experimental conditions allow for a non-reactive equimolar flow. We first consider a numerical idealization of this experiment on a square domain with $H_2$ and $N_2$ on the right half of the domain and $CO_2$ and $N_2$ on the left half of the domain. We employ the reduced formulation \cref{eq:MS2_reduced} with $c_1$ and $c_2$ denoting the $H_2$ and $N_2$ molar concentrations, respectively. Let $\Omega = [-0.5, 0.5]^2 \in \mathbb{R}^2$ with following initial conditions: 
\begin{align}
    c_{1}(x,y,t=0) &= 0.5/(1 + \exp(-20x)) \nonumber\\
    c_{2}(x,y,t=0) &= 0.5 \label{eq:IBVP1_duncan_n_toor}
\end{align}
We consider the binary diffusion coefficients provided by Boudin et.al.\cite{boudin2012mathematical}: $\mathcal{D}_{12} = 0.833$, $\mathcal{D}_{13} = 0.680$ and $\mathcal{D}_{23} = 0.168$. The simulation results for the molar concentrations are presented in \cref{fig:ms_duncan}, and the gradients of the molar concentrations are presented in \cref{fig:ms_duncan_grad}, where only the horizontal components of the gradients are presented as the fields are uniform 
in the vertical direction. We observe that the molar concentrations of $H_2$ and $CO_2$ behave in a familiar way, where diffusion has an equalizing effect of removing gradients in the molar concentration (see rows 1 and 3 of \cref{fig:ms_duncan_grad}) that results in uniform equilibrium steady state (see rows 1 and 3 of \cref{fig:ms_duncan}). However, $N_2$ shows an anomalous behavior. We first focus on the x-concentration gradient of $N_2$, \cref{fig:ms_duncan_grad}) and observe that there is no gradient initially. However, it evolves to form a peak at $t=0.25$, and then this peak decays to 0 at t=2. In terms of the concentration, $c_{N_2}$, this phenomenon appears as starting from a uniform field, $N_2$ first flows to the right, creating a gradient field and then flows back towards the left to a uniform equilibrium. This phenomenon we simulated through MS-diffusion cannot be explained through Fickian diffusion. In particular, at $t=0$, $N_2$ starts to redistribute even though  $\nabla c_{N2}=0$. This phenomenon is dubbed as osmotic diffusion that, in the case of Fickian diffusion, demands $D_{N_2}\rightarrow \infty$. In $t\in (0,0.25)$, the diffusion of $N_2$ is in a direction opposite to that dictated by its driving force in a Fickan model; 
this is known as reverse diffusion and requires $D_{N_2}<0$. At $t=0.25$, the diffusion mechanism switches from reverse diffusion to the usual Fickian diffusion. At this time, we observe a non-zero gradient field ($\nabla c_{N_2}\neq 0$); however, $J_{N_2}=0$. 
In Fickian diffusion, this will require $D_{N_2}=0$. This example shows that MS-diffusion can capture a richer set of diffusive mechanisms. In \cref{sec:interpretation_pattern}, we will discuss the relevance of these mechanisms on pattern formation. 

\section{Variational System Inference}\label{sec:vsi}

In this section, we will develop the framework for data-driven inference of the parameterized parabolic PDE to obtain the coefficients of MS diffusion and functional forms of production rates. We will first consider the variational formulation of \cref{eq:reaction_diffusion}. As discussed above, the equivalent diffusivity coefficients, $D_{ij}$, for MS diffusion, are nonlinear functions of the molar concentration $(c_1, \cdots c_{n-1})$. We will start with the general nonlinear form of diffusion coefficients and reaction terms. In \cref{sec:pde_infer}, we will assume a parameterized form for the production rate. 


\subsection{Finite element formulation for RD model}
We will consider $\bc\in \boldsymbol{\mathcal{V}}(\Omega)\times L_2([0,T])$ where the hilbert space, $\boldsymbol{\mathcal{V}}(\Omega) = H^{1}(\Omega)\times_{\text{n-1 times}}\cdots \times H^{1}(\Omega)$. For the infinite-dimensional initial and boundary value problem (IBVP) with homogeneous Neumann boundary conditions on $\partial \Omega$, the weak form can be stated as: for all admissible test functions $ \boldsymbol{w} \equiv (w_1, \cdots w_{n-1})\in \boldsymbol{\mathcal{V}}$ find $\boldsymbol{c}(\boldsymbol{x},t)$ such that $\boldsymbol{c}(\boldsymbol{x},t=0) = \overline{\boldsymbol{c}}(\boldsymbol{x})$ and for all $i\in \{1,\cdots {n-1}\}$

\begin{align}
    \int_\Omega \frac{\partial c_i}{\partial t} w_i d\Omega &=-\sum_j\int_\Omega     D_{ij} \nabla w_i \cdot \nabla c_jd\Omega  + \int_\Omega r_i w_i d\Omega 
\end{align}

\subsubsection{Time discretization}
For a finite time evolution $t\in [0,T]$, we consider finite timepoints $[t_0, t_1, \cdots, t_{n_T}]$ with $t_0 = 0$, $t_{n_T} = T$ and $t_k < t_{k+1}$. The timestep size is evaluated as $\Delta t_k = t_{k+1}-t_k$. We consider the backward Euler scheme for updating $c_i$

\begin{align}\label{eq:weak_time_discrete}
    \int_\Omega \frac{ c_i - c_i^{t_k}}{\Delta t_k} w_i d\Omega &=-\sum_j\int_\Omega     D_{ij} \nabla w_i \cdot \nabla c_jd\Omega  + \int_\Omega r_i w_i d\Omega 
\end{align}

\subsubsection{Finite element approximation}

For finite-dimensional fields $\boldsymbol{c}^h,\boldsymbol{w}^h$ the weak form is as follows: find $\boldsymbol{c}^h \in \boldsymbol{\mathcal{V}}^h \subset \boldsymbol{\mathcal{V}}$  such that for all $\boldsymbol{w}^h \in \boldsymbol{\mathcal{V}}^h \subset \boldsymbol{\mathcal{V}}$, the finite-dimensional (Galerkin) weak form of the problem is satisfied. The variation components $w_i^h$ and trial solutions $c_i^h$ are defined using a finite number of basis functions,
\begin{align}
    c_i^{h}  &= \sum_{k=1}^{n_b} a_{i,k} N_k\nonumber\\
    w_i^{h}  &= \sum_{k=1}^{n_b} b_{i,k} N_k
\end{align}
where $n_b$ is the dimensionality of the $H^{1,h}(\Omega)$, and and $N_k$ represents the global basis functions. Substituting $w^h$ and $c^h$ in \cref{eq:weak_time_discrete}, accounting for arbitrariness of $w_i^{h}$, and decomposing the integration over $\Omega$ by a sum over subdomains $\Omega_e$ leads to following set of residual equations at each timestep for $i\in \{1, \cdots {n-1}\}$ and $l\in \{1, \cdots n_b\}$:
\begin{align}
    R_{i,l,s} = \sum_{e=1}^{n_e} \sum_{k=1}^{n_{b_e}} \int_{\Omega_e}  \left( \frac{a_{i,e(k)} - a_{i,e(k)}^{t_{s-1}}}{\Delta t_s} N_k N_l d\Omega 
    + \sum_jD_{ij} a_{j,e(k)}\nabla N_{e(k)} \cdot \nabla N_{l} 
    - r_i N_l\right) d\Omega
\end{align}
The solution to the PDE at each timestep is obtained by solving the above set of $n\times n_b$ equations, $R_{i,l} = 0$ for all $i$ and $l$. Note that these equations are nonlinear in general as $D_{ij}$ and $r_j$ are typically functions of $\boldsymbol{c}$. 

\subsection{Identification of governing parabolic PDEs in weak form}\label{sec:pde_infer}
 
We are provided with the data as a spatiotemporal field for the multicomponent system, $\bc(\bx,t)$ with $(\bx, t) \in \Omega \times [0,T]$ where $\Omega \subset \mathbb{R}^n$ and $[0,T]$ are the spatial domain and time-period of interest. We will assume that irrespective of the source of data, be it experimental images or high-fidelity simulations, it is possible to obtain the finite element projection of the data field as $c^{d,h}_i = \sum_k a^d_{i,k} N_k$ where we are denoting the data-field as $\bc^d$. 

The exact parametric form of PDE can be obtained by considering a set of possible reaction operators. In this work, we consider polynomial operators for reactive terms. 
\begin{equation}
    r_i (\boldsymbol{c}; \boldsymbol{\theta}) = \sum_{k=0}^{n_o} \sum_{|\boldsymbol{\alpha}|=k} \theta_{i,\alpha}\boldsymbol{c}^{\boldsymbol{\alpha}}
\end{equation}
where the multi-index notation with $\boldsymbol{\alpha}\equiv (\alpha_1, \cdots, \alpha_{n-1})$ with  $|\boldsymbol{\alpha}| = \alpha_1 + \cdots + \alpha_{n-1}$, and $\boldsymbol{c}^{\boldsymbol{\alpha}} = c_1^{\alpha_1} c_2^{\alpha_2}\cdots c_{n-1}^{\alpha_{n-1}}$. And, $n_o$ defines some cut-off order for the polynomial approximation. 

These polynomial forms are a popular choice in modeling pattern-forming RD models such as Schnackenberg, Gray Scott and FitzHugh-Nagumo models. It should also be noted that the \textit{no-net production} is implicitly built in this ansatz. This can be observed by fixing the reactive dynamics of the $n^{\text{th}}$ component as $r_n (\boldsymbol{c}; \boldsymbol{\theta}) = \sum_{k=0}^{n_o} \sum_{|\boldsymbol{\alpha}|=k} \theta_{n,\alpha}\boldsymbol{c}^{\boldsymbol{\alpha}}$ with $\theta_{n,\alpha} = -\sum_{i=1}^{n-1}\theta_{i,\alpha}$. This results in the required condition that $\sum_{i=1}^{n} r_i = 0$.
The condition for \textit{positivity-preserving} is more challenging to achieve. A sufficient condition for this property to hold is that we only consider polynomials of the form $\boldsymbol{c}^{(1,1,\cdots, 1)+\boldsymbol{\alpha}}$ forcing $r_i(\boldsymbol{c}, \boldsymbol{\theta})=0$ if any of the components are zero. However, this is a very restrictive condition. Since it is sufficient and not necessary, we will not enforce \textit{positivity-preserving} property in this work and restrict our attention to models and respective IbVPs that naturally remain in the positive quadrant.

With this ansatz in place, we can formulate the optimization problem for PDE estimation. The set of optimization variables are given as $\boldsymbol{\Theta} = \{ \mathcal{D}_{ij}\}_{i\neq j\in{1, \cdot n}} \cup \{\theta_{i,\boldsymbol{\alpha}}\}_{i\in \{1, \cdot n-1\}, |\boldsymbol{\alpha}|\leq n_o}$. The optimization problem is posed as: 
\begin{align}\label{eq:VSI_loss}
    \boldsymbol{\theta}^* = \argmin_{\boldsymbol{\theta} \in \boldsymbol{\Theta}} \left\vert\left\vert \bR(\boldsymbol{\theta}| \bc^d) \right\vert\right\vert^2, \quad \text{ such that } \mathcal{D}_{ij}\geq 0 \quad\forall i,j
\end{align}
where $\bR(\boldsymbol{\theta}| \bc^d)$ is an $(n-1) \times n_b \times n_T$-dimensional vector indexed as $R_{i,l,s}$ with $(i, l, s) \in \{1, \cdots n-1,\}\times \{1, \cdots n_b\}\times \{1, \cdots n_T\}$ where: 
\begin{align}
    R_{i,l,s} (\boldsymbol{\theta|\boldsymbol{c}^d}) = \sum_e \sum_{k=1}^{n_b} \int_{\Omega_e}  \left( \frac{a_{i,k}^{d,t_{s}} - a_{i,k}^{d,t_{s-1}}}{\Delta t_s} N_k N_l d\Omega 
    + \sum_jD_{ij}(\boldsymbol{c}^d; \boldsymbol{\theta}) a_{j,k}^{d,t_{s}}\nabla N_{k} \cdot \nabla N_{l} 
    - r_i(\boldsymbol{c}^d; \boldsymbol{\theta}) N_l\right) d\Omega
\end{align}

\subsubsection{Inference with MS diffusion}

We first employ this VSI approach on MS diffusion. In particular, we explore the equimolar non-reactive transport example of Duncan and Toor's experiment (see \cref{eq:IBVP1_duncan_n_toor}). From here on, we will refer to $H_2$, $N_2$ and $CO_2$ as species 1, 2, and 3, respectively. We are interested in inferring reduced 2-component MS equation \cref{eq:MS2diffusion} parameterized with $\boldsymbol{\Theta} = \{\mathcal{D}_{12}, \mathcal{D}_{23}, \mathcal{D}_{13} \}$. We consider 3 different IBVPs with IBVP1 same as \cref{sec:dunan_n_toor} and IBVP2 and 3 defined as:
\begin{align}
    \text{IBVP 2}:\quad c_{1}(x,y,t=0) &= 0.25 + 0.1\cos(10x) + 0.1\cos(10y) \nonumber\\
    c_{2}(x,y,t=0) &= 0.25 + 0.1\sin(10x) + 0.1\sin(10y)\nonumber\\
    \text{IBVP 3}:\quad c_{1}(x,y,t=0) &= 0.5/(1 + \exp(-20x)) \nonumber\\
    c_{2}(x,y,t=0) &= 0.5/(1 + \exp(20x))   \label{eq:IBVP23_duncan_n_toor}
\end{align}

\begin{figure}
    \centering
    \includegraphics[width=0.95\linewidth]{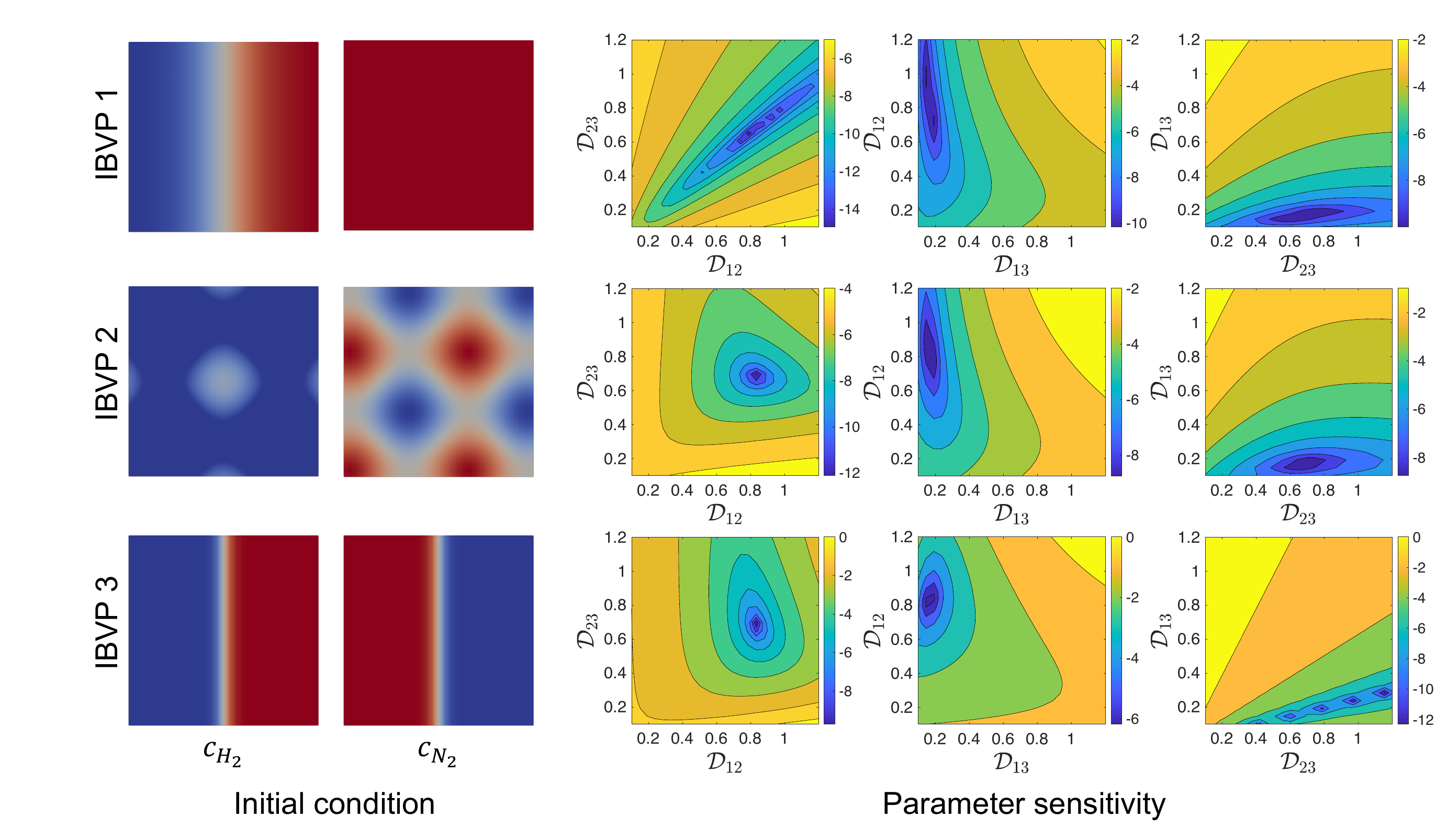}
    \caption{Parameter sensitivity for different IBVPs. Each row represent a different IBVP with the exact form provided in \cref{eq:IBVP1_duncan_n_toor,eq:IBVP23_duncan_n_toor}. The left panel shows the initial concentration of the dataset. The right panel shows sensitivity contours (left to right) in the $(\mathcal{D}_{23},\mathcal{D}_{12})$, $(\mathcal{D}_{13},\mathcal{D}_{12})$, $(\mathcal{D}_{23},\mathcal{D}_{13})$ planes while fixing $\mathcal{D}_{13}$, $\mathcal{D}_{23}$ and $\mathcal{D}_{12}$, respectively to the original solution. The sensitivity contours are constructed using $\log ||\bR(\boldsymbol{c}^d| \boldsymbol{\theta})||$}
    \label{fig:ms2_contour}
\end{figure}

\begin{table}[]
\begin{tabular}{|c|l|l|l|l|}
\hline
Parameter & Original & Inferred (IBVP 1) & Inferred (IBVP 2)& Inferred (IBVP 3) \\ \hline
$\mathcal{D}_{12}$ &   0.833    &0.83298162& 0.83305782& 0.83300214\\
$\mathcal{D}_{23}$ &   0.680    &0.68001752& 0.67998865 &0.24130419     \\
$\mathcal{D}_{13}$ &   0.168    &0.16799677& 0.1679995& 0.05961649     \\ \hline
\end{tabular}
\caption{Inferred parameters for a 2-component MS diffusion with different IBVPs}
\label{tab:MS2}
\end{table}
The IBVPs are also provided in \cref{fig:ms2_contour} (left panel). 
We infer the parameters for 2-component MS diffusion by solving \cref{eq:VSI_loss}, and the inferred parameters are presented in \cref{tab:MS2}. We observe that for the case of IBVP 1 and IBVP 2, we can recover the PDE to a reasonable precision. However, that is not the case in IBVP 3. 
We can also consider the sensitivity of the residual to the parameters by evaluating the contours for a range of parameter space (shown in the right panel \cref{fig:ms2_contour}). We can observe from the bottom-right contour plot $(\mathcal{D}_{23},\mathcal{D}_{13})$ for IBVP 3 that multiple minima are present along a line. This results in a degeneracy in the inferred parameters due to the chosen initial conditions. One may notice that in IBVP 3, $c_1 + c_2 \approx 1$ almost everywhere. It follows that: 
\begin{itemize}
\item For a ternary system, if $c_1 + c_2 = 1 $ at any given point in time then $\frac{\partial (c_1 + c_2)}{\partial t} =0$ i.e. if $c_1 + c_2 =1 $ at some time, then it remains $1$ for all times. Consequently there is no information of interactions of the third species, resulting in degeneracy in the inferred frictional forces that involves this species,
\item If $\boldsymbol{c}^d$ is such that it satisfies the above condition and $(\mathcal{D}_{12}^*, \mathcal{D}_{23}^*, \mathcal{D}_{13}^*)$ are such that \cref{eq:MS2_reduced} is satisfied, then choosing $(\mathcal{D}_{12}^*, \beta \mathcal{D}_{23}^*, \beta  \mathcal{D}_{13}^*)$ for any will also satisfy \cref{eq:MS2_reduced}. 
\end{itemize}

The first point is easy to verify by adding the two components of \cref{eq:MS2_reduced}: 
\begin{align*}
    \frac{\partial{(c_1 + c_2)}}{\partial t} = \nabla.\left( \frac{1-c_1-c_2}{\det \bB} \left( \frac{\nabla c_1}{\mathcal{D}_{13}} - \frac{\nabla c_2}{\mathcal{D}_{23}}\right) + \frac{1}{\mathcal{D}_{12}\det \bB} \nabla (c_1 + c_2)^2
    \right)
\end{align*}
Noting that $c_1 + c_2 = c_{\text{tot}} = 1$ and $\nabla(c_1 + c_2)^2 = \nabla c_{\text{tot}}^2 = 1$, we get that $\frac{\partial{(c_1 + c_2)}}{\partial t} = 0$

To show the second point, we first observe that $\frac{\partial{c_1}}{\partial t} = - \frac{\partial{c_2}}{\partial t} $, therefore it is only necessary to consider evolution of one of the species. Without loss of generality, we consider the evolution of $c_1$ 
\begin{align*}
    \frac{\partial c_1}{\partial t} = \nabla \cdot \left( \frac{1}{\det \bB} \left( \left(\frac{1}{\mathcal{D}_{13}} - \frac{1}{\mathcal{D}_{23}}\right) \nabla c_1^2 + \frac{\nabla c_1}{\mathcal{D}_{23} } \right) \right)
\end{align*}
here we eliminated $c_2$ using the additional requirement $ c_2 = 1 - c_1$. If the data, $\bc^d$ satisfies this property, one can write the following inference consistency error, $e(\mathcal{D}_{12}, \mathcal{D}_{23}, \mathcal{D}_{13}|c_1^d)$ for the data $c_1^d$ : 
\begin{align*}
    e(\mathcal{D}_{12}, \mathcal{D}_{23}, \mathcal{D}_{13}| c_1^d) = \left|\left|     \frac{\partial c_1^d}{\partial t} - \nabla \cdot \left( \frac{1}{\det \bB (\mathcal{D}_{12}, \mathcal{D}_{23}, \mathcal{D}_{13}, c_1^d)} \left( \left(\frac{1}{\mathcal{D}_{13}} - \frac{1}{\mathcal{D}_{23}}\right) \nabla (c_1^{d})^2+ \frac{\nabla c_1^d}{\mathcal{D}_{23} } \right) \right) \right|\right|
\end{align*}
where \begin{align*}
    \det \bB = \frac{1-c_1}{\mathcal{D}_{12}\mathcal{D}_{23}}- \frac{c_1}{\mathcal{D}_{13}\mathcal{D}_{12}}
\end{align*}
Here the inference consistency error, $e$, provides a measure of how closely the data field, $c_1^d$ and the parameters $(\mathcal{D}_{12}, \mathcal{D}_{23}, \mathcal{D}_{13})$ correspond to each other. In particular, if $(\mathcal{D}_{12}^*, \mathcal{D}_{23}^*, \mathcal{D}_{13}^*)$ is the parameter that generated the solution $c_1^d$ then $e(\mathcal{D}_{12}^*, \mathcal{D}_{23}^*, \mathcal{D}_{13}^* | c_1^d)) = 0$. We also observe that for any $\beta>0$: 
\begin{align*}
    e(\mathcal{D}_{12}, \beta\mathcal{D}_{23}, \beta\mathcal{D}_{13}| c_1^d) = 
    \left|\left|     \frac{\partial c_1^d}{\partial t} - \nabla \cdot \left( \frac{1}{\det \bB (\mathcal{D}_{12}, \beta\mathcal{D}_{23}, \beta\mathcal{D}_{13}, c_1^d)} \left( \left(\frac{1}{\beta \mathcal{D}_{13}^*} - \frac{1}{\beta \mathcal{D}_{23}^*}\right) \nabla (c_1^{d})^2+ \frac{\nabla c_1^d}{\beta \mathcal{D}_{23}^* } \right) \right) \right|\right| \\
    =     \left|\left|     \frac{\partial c_1^d}{\partial t} - \nabla \cdot \left( \frac{1}{\det \bB (\mathcal{D}_{12}, \mathcal{D}_{23}, \mathcal{D}_{13}, c_1^d)} \left( \left(\frac{1}{ \mathcal{D}_{13}^*} - \frac{1}{ \mathcal{D}_{23}^*}\right) \nabla (c_1^{d})^2+ \frac{\nabla c_1^d}{ \mathcal{D}_{23}^* } \right) \right) \right|\right|=e(D_{12}^*, D_{23}^*, D_{13}^*) 
    =0
\end{align*}
 
This degeneracy can be generalized to n-component systems where if any of the components is uniformly zero at any given time, then the binary coefficients corresponding to that component are degenerate. Using these arguments for the three IBVPs, we see that this degeneracy is absent in IBVPs 1 and 2 since we start with a mixture with a non-zero concentration of each component. Therefore, the inference is more robust for these two cases. 


\subsection{Computational framework for inference}
The numerical examples presented in this work
have been posed and solved in two dimensions using the finite element method programmed using the automated computational framework of \textit{Fenics} package\cite{dolfin, logg2012automated}. All the simulations presented in this work are done on a 2D square domain with structured mesh with 5000 linear triangular elements. The nonlinear variational problem in the forward solution is solved using a Newton solver. The nonlinear optimization for minimizing \cref{eq:VSI_loss} is done using \textit{L-BFGS-B} method. We use the implementation provided by the \textit{scikit-learn} package\cite{scikit-learn}. The objective function,     $\texttt{obj} = \left\vert\left\vert \bR(\boldsymbol{\theta}| \bc^d) \right\vert\right\vert^2$ is evaluated using the \textit{Fenics} library using the same mesh that the data was generated on. The gradient of the objective function is also supplemented to the optimization scheme, where it is evaluated as $\texttt{grad\_obj} = \left\vert\left\vert \bR(\boldsymbol{\theta}| \bc^d) \right\vert\right\vert \nabla_{\boldsymbol{\theta}}\bR(\boldsymbol{\theta}| \bc^d) $, where the gradient of the residual is estimated using the automatic differentiation capability of \textit{Fenics} package.


\section{Pattern formation in RD model}\label{sec:pattern}

Turing's work showed that two or more interacting components diffusing at significantly different rates can create spontaneous, self-organizing spatial concentration patterns. In this section, we first present the analysis for MS-based RD systems, followed by a discussion on the inference of these systems. However, the ideas presented here apply to the general n-component model. We illustrate these properties for a ternary mixture represented with the concentration of 2-components, $(c_1, c_2)$, with dynamics governed by an RD system composed of MS diffusion and Schnackenberg kinetics for reaction, resulting in the following set of equations: 
\begin{align}
    \frac{\partial c_1}{\partial t} &= \nabla \cdot D_{11}(c_1, c_2) \nabla c_1 + \nabla \cdot D_{12}(c_1, c_2) \nabla c_2 + R_{10} + R_{11} c_1 + R_{12} c_1^2 c_2   \nonumber\\
    \frac{\partial c_2}{\partial t} &= \nabla \cdot D_{21}(c_1, c_2) \nabla c_1 + \nabla \cdot D_{22}(c_1, c_2) \nabla c_2 + R_{20} + R_{21} c_1 + R_{22} c_1^2 c_2   \label{eq:schnackenberg}
\end{align}
Here, the terms $D_{ij}(c_1, c_2) = (\bB^{-1})_{ij}$ represent the nonlinear diffusion obtained by solving \cref{eq:MS2diffusion}. The reactive dynamics represented by $r_1 = R_{10} + R_{11} c_1 + R_{12} c_1^2 c_2$, and $r_2 = R_{20} + R_{21} c_1 + R_{22} c_1^2 c_2$ is known as the Schnackenberg dynamics \cite{garikipati2017perspectives} and is the basis for many biological patterns. This system contains 3 parameters for diffusion, namely $(\mathcal{D}_{12}, \mathcal{D}_{23}, \mathcal{D}_{13})$, and 6 parameters for reaction, namely $(R_{10}, R_{11}, R_{12}, R_{20}, R_{21}, R_{22})$. It is also worth noting that the dynamics of the concentration of the $3^{\text{rd}}$ species can be determined using the relationship: 
\begin{align}
    \frac{\partial c_3}{\partial t} = - 
    \frac{\partial c_1}{\partial t}  - 
    \frac{\partial c_2}{\partial t}, \qquad c_3(\bx, 0) = 1-c_1(\bx, 0) - c_2(\bx, 0)  
\end{align}

\subsection{Turing's analysis with nonlinear diffusion}

\begin{figure}
    \centering
    \includegraphics[width=0.95\linewidth]{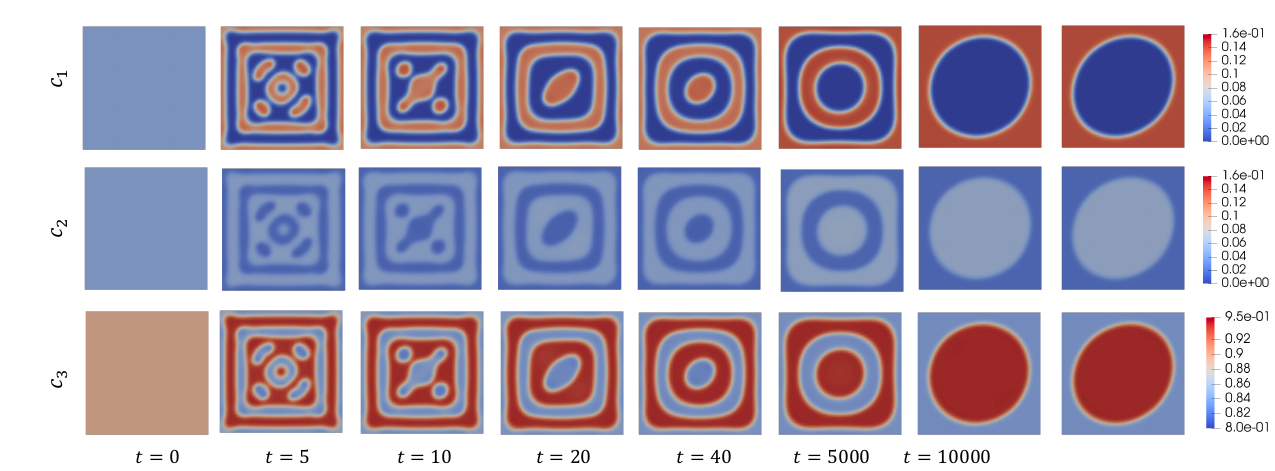}
    \caption{A numerical solution of \cref{eq:schnackenberg} on $\Omega = [-5,5]^2$ with parameters $(\mathcal{D}_{12}, \mathcal{D}_{13}, \mathcal{D}_{23}) = (0.833, 0.680,0.168)$ and $(R_{10}, R_{11}, R_{12}, R_{20}, R_{21}, R_{22}) = (0, -20, 7200, 0, 20, -7200)$. The simulation is started with the unstable equilibrium solution $\bc(\bx, 0)=(0.0527, 0.0527)$ with a perturbation of $10^{-4}\sin(||x||)/(1e-3+||x||)$ }
    \label{fig:forward_schnackenberg}
\end{figure}

In this section, we extend Turing's analysis to MS-based RD systems. Turing instability is a linear instability that is observed in some RD systems for specific choices of parameters. To observe this phenomenon, one starts with a homogeneous IBVP, i.e. $\bc(\bx,0)\equiv \bc_0$ is a constant vector in space. This constant vector $\bc_0$ is selected such that $r_i(\bc)=0$ for all $i\in \{1,\cdots, n\}$, i.e. the initial homogeneous condition is a solution to the system of nonlinear reaction equations. As the solution is homogeneous, \cref{eq:MS2flux_reduced} results in the absence of diffusive flux at all points in the domain. Consequently, this homogeneous solution is also a solution to the full RD system, and one should expect that if this homogeneous solution is attained at any time during evolution, then the system will remain at this solution for all future times. However, for certain choices of parameters, if this solution is perturbed from its equilibrium, these perturbations may grow, causing the system to evolve to a non-homogeneous equilibrium state. A typical observation of this phenomenon is presented in \cref{fig:forward_schnackenberg} where a numerical solution to \cref{eq:schnackenberg} is presented on $\Omega = [-5,5]^2$ with parameters $(\mathcal{D}_{12}, \mathcal{D}_{13}, \mathcal{D}_{23}) = (0.833, 0.680,0.168)$ and $(R_{10}, R_{11}, R_{12}, R_{20}, R_{21}, R_{22}) = (0, -20, 7200, 0, 20, -7200)$. The simulation is started at the homogeneous state $\bc_0=(0.0527, 0.0527)$ with a perturbation of $10^{-4}\sin(||x||)/(1e-3+||x||)$. We can observe that this homogeneous state is a solution to the Schnackenberg equation for the prescribed parameters. Moreover, the magnitude of perturbation is small in comparison to the homogeneous solution. In the simulations, we first focus on the evolution of $c_1$ and $c_2$, noting that the fields rapidly move towards a non-homogeneous solution in the first few time steps, forming concentric rings of localized high and low concentrations. The changes in these concentration fields become slower as time progresses, and beyond a certain time, there is no discernible change in these concentration fields. The inhomogeneous nature of these equilibrium fields is what we refer to as patterns in RD systems. Turing\cite{turing1990chemical} explained the onset of this instability that we observed at $t=0$ using the stability of the linearized RD system. 



Noting that the initial condition is a homogeneous equilibrium solution with some added perturbation, we consider the following general form of perturbed solution about an equilibrium point: 
\begin{align*}
    c_i(\bx) = \overline{c}_i + \widetilde{c}_i(\bx)
\end{align*}
where $\overline{\bc} = (\overline{c}_1, \cdots, \overline{c}_{n-1)}$ represent the equilibrium solution and the $\widetilde{c}_i$ represent the perturbations about the $i^{\text{th}}$ component equilibrium solution. Substituting this form in \cref{eq:reaction_diffusion} gives
\begin{equation}\label{eq:linearizedRD_general}
\frac{\partial \widetilde{c}_i}{\partial t} = \sum_{j} D_{ij}(\overline{\bc}) \Delta \widetilde{c}_j +  \frac{\partial r_i}{\partial c_j} (\overline{\bc}) \widetilde{c}_j
\end{equation}
where we used the fact that $\overline{\bc}$ is an equilibrium point i.e. for all $i\in\{1,\cdots, n-1\}$, we have $\frac{\partial \overline{c}_i}{\partial t} = \sum_{j} D_{ij}(\overline{\bc}) \Delta \overline{c}_j +  r_i(\overline{\bc}) $. Moreover, the homogeneous solution is gradient-free, i.e. for all $i\in\{1,\cdots, n-1\}$ and $i\in\{1,\cdots, n_d\}$, we have  $\frac{\partial \overline{c}_j}{\partial x_i} = 0$. \cref{eq:linearizedRD_general} approximates the dynamics of perturbation growth when the solution is \textit{close} to a homogeneous equilibrium solution. For the special case of Schnackenberg dynamics, this equation gives: 

\begin{align}\label{eq:linearSchnack}
    \frac{\partial \widetilde{c}_1}{\partial t} &= \overline{D}_{11}
 \Delta \widetilde{c}_1 + \overline{D}_{12}
 \Delta \widetilde{c}_2 + \overline{R}_{11} \widetilde{c}_1 + \overline{R}_{12}  \widetilde{c}_2   \nonumber\\
    \frac{\partial \widetilde{c}_2}{\partial t} &= \overline{D}_{21}
 \Delta \widetilde{c}_1 + \overline{D}_{22}
 \Delta \widetilde{c}_2 + \overline{R}_{21} \widetilde{c}_1 + \overline{R}_{22}  \widetilde{c}_2   
\end{align}

where:
\begin{align*}
    \overline{R}_{11} &= R_{11} + 2R_{12} \overline{c}_1 \overline{c}_2 \\
     \overline{R}_{12} &= R_{12} \overline{c}_1^2 \\
     \overline{R}_{21} &= R_{21} + 2R_{22} \overline{c}_1 \overline{c}_2 \\
     \overline{R}_{22} &= R_{22} \overline{c}_1^2 
\end{align*}

We consider solutions of the following form: 
\begin{align}\label{eq:omega_growth}
    \widetilde{c}_{\alpha} = \sum_{k} T_{\alpha,k} e^{i\boldsymbol{k}\cdot\boldsymbol{x}} e^{\omega_k t}
\end{align}
with wave vector $\boldsymbol{k}$ and frequency $\omega_k$ for the plane wave mode $k$. Substituting in \cref{eq:linearSchnack}: 
\begin{align*}
    \left( |\boldsymbol{k}|^2 \overline{D}_{11} - \overline{R}_{11} +\omega_k\right)  T_{1,k} + \left( |\boldsymbol{k}|^2 \overline{D}_{12} - \overline{R}_{12} \right)T_{2,k} =0 \nonumber\\
    \left( |\boldsymbol{k}|^2 \overline{D}_{21} - \overline{R}_{21} \right)  T_{1,k} + \left( |\boldsymbol{k}|^2 \overline{D}_{22} - \overline{R}_{22} +\omega_k \right)T_{2,k} =0
\end{align*}
For non-trivial $T_{1,k}$ and $T_{2,k}$, the frequency and wave number must satisfy: 
\begin{align}\label{eq:freq_mode}
    0 = w_k^2 +  b\omega_k + c  
\end{align}

where
\begin{align*}
    b &=    |\boldsymbol{k}|^2 (\overline{D}_{11}+ \overline{D}_{22}) - (\overline{R}_{11} + \overline{R}_{22}) \\
    c &= \left( \left( |\boldsymbol{k}|^2 \overline{D}_{11} - \overline{R}_{11} \right)\left( |\boldsymbol{k}|^2 \overline{D}_{22} - \overline{R}_{22}\right) - \left( |\boldsymbol{k}|^2 \overline{D}_{12} -  \overline{R}_{12} \right)\left( |\boldsymbol{k}|^2 \overline{D}_{21} - \overline{R}_{21} \right) \right)  
\end{align*}
This quadratic equation has the solutions, $\omega_k = 1/2(-b\pm \sqrt{b^2-4c})$. We will refer to these solutions as the branches. The form in \cref{eq:omega_growth} reveals that
if $\text{Real}(\omega_k)>0$ for any value of $k$ and any branch, then the corresponding mode will grow over time. This growth is typically expected only close to the homogeneous equilibrium solution, $\overline{\bc}$. As these modes grow, the state departs from the homogeneous solution and may stabilize at non-homogeneous equilibrium points that we refer to as \textit{`patterns'}. 
One way of studying this growth in the linearized regime is via stability diagrams. In this visualization, we plot $\text{Real}(\omega_k)$ for the different branches for a range of wave numbers $k$. If any region of the figure lies in the first quadrant, then the system is linearly unstable. Moreover, the values of $k$ for which $\text{Real}(\omega_k)>0$ determine the modes that will grow. In particular, one may identify the different modes in the perturbation and use this visualization to determine the unstable ones. The unstable branch of the system discussed in \cref{fig:forward_schnackenberg} is presented in \cref{fig:time_training_stability}. In this case, all modes are observed to be unstable. In another case study presented in \cref{sec:inference_pattern}, we get a stability diagram (shown in \cref{fig:case1_pattern}) where not all modes are unstable. It should be remarked that the patterns are formed when the system's state evolves beyond a small neighborhood of homogeneous equilibrium solution. Therefore, this stability analysis is only valid at the onset of the linear instability and not at all subsequent times. 
It should also be observed that different nonlinear systems can have the same onset to instability, i.e. two different systems may have the same instability diagram. For instance, consider the following two RD systems with the same linearized reactions but different diffusivities.  
\begin{align}
 \text{Case 1}: &\quad \begin{bmatrix}
    \overline{R}_{11}  & \overline{R}_{12} \\
    \overline{R}_{21} & \overline{R}_{22}
\end{bmatrix}= \begin{bmatrix}
    -1 & 1 \\
    -1 & 1
\end{bmatrix} , \quad    \begin{bmatrix}
    \overline{D}_{11}  & \overline{D}_{12} \\
    \overline{D}_{21} & \overline{D}_{22}
\end{bmatrix}= \begin{bmatrix}
    1 & 0 \\
    0 & 1/100
\end{bmatrix} \nonumber\\
 \text{Case 2}: &\quad \begin{bmatrix}
    \overline{R}_{11}  & \overline{R}_{12} \\
    \overline{R}_{21} & \overline{R}_{22}
\end{bmatrix}= \begin{bmatrix}
    -1 & 1 \\
    -1 & 1
\end{bmatrix} ,
\quad \begin{bmatrix}
    \overline{D}_{11}  & \overline{D}_{12} \\
    \overline{D}_{21} & \overline{D}_{22}
\end{bmatrix} =  \begin{bmatrix}
    505/1000 & 99 (\sqrt{2}+1)/200 \\
    99 (\sqrt{2} - 1)/200 & 505/1000
\end{bmatrix} \label{eq:example_linRD}
\end{align}
Solving \cref{eq:freq_mode} for both these cases results in the same frequency-wave number relationship:
\begin{align*}
        \omega_k = \frac{1}{2} \left( - \frac{101}{50}|\boldsymbol{k}|^2  \pm \sqrt{\left(2+\frac{99}{100}|\boldsymbol{k}|^2 \right)^2 -4} \right)
\end{align*}
Therefore, these two cases behave similarly at the homogeneous equilibrium state. 
It is also worth noting that in case 1, the components are diffusing at significantly different rates; meanwhile, in case 2, the instability arises due to cross-diffusion terms. Regarding physical mechanisms of diffusion, these cases represent two entirely different 
approaches to the onset of pattern formation. These mechanisms have been previously studied in literature\cite{madzvamuse2015cross,vanag2009cross} in the context of Fickian diffusion. In the next section, we will discuss the physical mechanisms in the MS-based RD model.

\subsubsection{Physical interpretation of Turing instability in MS-based RD systems}\label{sec:interpretation_pattern}

Turing instabilities were initially studied for a Fickian diffusion-based RD system\cite{turing1990chemical}. In this case, diffusivities $D_{ij}$ are independent of concentrations. The simplicity of the Fick-based RD system allows for an activator-inhibitor interpretation of the morphogens that lead to instability and, consequently, steady-state patterns. To understand this interpretation, we can consider the linearized system in \cref{eq:linearSchnack}. For this brief excursion into Fick-based RD systems, we will ignore the species, $c_3$, that in the case of MS-based diffusion is determined as $c_3(c_1, c_2) = 1-c_1-c_2$ (conservation of total molar concentration). Moreover, we will ignore the dependence of linearized coefficients $\overline{D}_{ij}$ and $\overline{R}_{ij}$ on the equilibrium homogeneous molar concentrations, $(\overline{c}_1, \overline{c}_2)$. In this setting, we will refer to  $\overline{R}_{11}, \overline{R}_{22} > 0$ as auto-activation, $\overline{R}_{11}, \overline{R}_{22} < 0$ as auto-inhibition, $\overline{R}_{12}, \overline{R}_{21} > 0$ as cross-activation, and $\overline{R}_{12}, \overline{R}_{21} < 0$ as cross-inhibition. Moreover, for diffusion, thermodynamic admissibility requires that self-diffusion $\overline{D}_{11}, \overline{D}_{22}>0$ meanwhile cross-diffusion ($\overline{D}_{12}, \overline{D}_{21}$) can be positive or negative. One of the simplest cases with instability in this activator-inhibitor setting is in the absence of cross-diffusion ($\overline{D}_{12}= \overline{D}_{21} = 0$) with fast inhibitor (species 1) and slow activator (species 2) ($\overline{D}_{11}>> \overline{D}_{22} $). Here, inhibitors are species with auto and cross-inhibitions ($\overline{R}_{11},\overline{R}_{12}<0$) and activators are species with auto and cross-activations ($\overline{R}_{22},\overline{R}_{21}>0$). An example of this scenario is presented in case 1 of \cref{eq:example_linRD}. In this scenario, if there is a perturbation that results in a locally higher concentration of activator, the activator results in a local increase of the inhibitor. However, inhibitors having a faster spreading rate result in a long-range inhibition. Consequently, pushing the system towards a non-homogeneous solution with low and high concentrations. This process is often referred to as
short-range activation and long-range inhibition. It should also be noted that this onset to instability doesn't require different self-diffusions as in case 1 of \cref{eq:example_linRD}. It is shown in case 2 of \cref{eq:example_linRD} that the same $(\omega_k - \bk)$ relationship can be obtained with species with the same self-diffusion but different cross-diffusions. In this scenario, the different spreading rate in the activator-inhibitor model is achieved through this cross-diffusion mechanism.

It is evident from the above discussion that various diffusive mechanisms can lead to pattern formation. It is, therefore, important to have a physically accurate description of diffusion mechanisms in real-life systems. In \cref{sec:dunan_n_toor}, we considered Duncan and Toor's experiment that serves as a case study where Fickian diffusion is inadequate to describe transport in a ternary system. MS diffusion describes a more general class of diffusive mechanisms as discussed in \cref{sec:dunan_n_toor}. At the same time, MS-diffusion remains thermodynamically consistent\cite{KRISHNA1997} for positive coefficients, $\mathcal{D}_{ij}>0$. The form of diffusion provided in \cref{eq:MS2diffusion} shows a concentration-dependent form. It can be observed that at low concentrations i.e. $c_1,c_2<<1$, \cref{eq:MS2diffusion} gives:
\begin{equation*}
    \bB^{-1} \approx \begin{bmatrix}
    \mathcal{D}_{13} & 0 \\
    0 & \mathcal{D}_{23}
    \end{bmatrix}
\end{equation*}
Therefore, if the frictional forces between species (1-3) and species (2-3), given as $\mathcal{D}_{13}^{-1}$ and $\mathcal{D}_{23}^{-1}$ are very different, we may obtain the fast and slow diffusing species that sometimes form the basis for the activator-inhibitor model. Even at higher densities, the relative values of frictional forces are essential in understanding the diffusion mechanisms; for instance, at the onset of instability in the example presented in \cref{fig:forward_schnackenberg}, the coefficients of linearized RD system  \cref{eq:linearSchnack} evaluates to be:
\begin{align}
\begin{bmatrix}
    \overline{R}_{11}  & \overline{R}_{12} \\
    \overline{R}_{21} & \overline{R}_{22}
\end{bmatrix}= \begin{bmatrix}
    39.9860 & 19.9930 \\
    -39.9860 & -19.9930
\end{bmatrix} , \quad    \begin{bmatrix}
    \overline{D}_{11} & \overline{D}_{12} \\
    \overline{D}_{21} & \overline{D}_{22}
\end{bmatrix}= \begin{bmatrix}
    0.1755 & -0.0302 \\
    -0.0017 & 0.6869
\end{bmatrix}
\end{align}
Here, we observe that the self-diffusions are of similar magnitude while there is the presence of cross-diffusion terms. These examples show that MS-based RD systems can exhibit equivalent diffusion tensors as Fickian diffusion at the onset of instability. However, MS diffusion remains thermodynamically consistent for all molar concentrations in the permissible range. More interestingly, the concentration-dependent diffusion implies that a spatially and temporally varying diffusion is observed as concentration evolves after being perturbed from the unstable equilibrium. For instance, in some regions, the system may evolve with slow and fast diffusing species, while in others, it may evolve with cross-diffusion. This out-of-equilibrium dynamics is significantly more complex than Fick-based RD systems. On the other hand, this rich dynamics complicates the inverse problem of inferring parameters in MS-based RD systems due to the nonlinearities introduced by the diffusion process in the optimization problem of \cref{eq:VSI_loss}. 

\subsection{Data-driven inference of governing equations for pattern formation}\label{sec:inference_pattern}

In this section, we present two numerical studies to demonstrate the application of the data-driven system inference method outlined in \cref{sec:pde_infer} to pattern formation. The coefficients of the exact and inferred models are presented in \cref{tab:schnack}. 

\subsubsection{Data generation}
We consider two particular instances of the RD model with MS diffusion. The parameters are prescribed in Table \ref{tab:schnack}. In both cases, we infer a model by identifying the coefficients of the reaction basis terms as they appear in the Schnackenberg model, i.e. $\{c_1, c_2, c_1^2 c_2\}$ for both components. The models are inferred by solving the optimization problem of \cref{eq:VSI_loss}.\\
\textit{Case A:} The model with parameters labeled \textit{Exact A} is solved on the domain $\Omega = [-5,5]^2$ with $t\in[0,2]$. The initial conditions are chosen to be the perturbed unstable equilibrium as $\boldsymbol{c}(\boldsymbol{x},0) = (0.055,0.046)+\boldsymbol{\delta}$ where the perturbation, $\boldsymbol{\delta} = 0.01(\cos(10x)+\cos(10y), \sin(10x)+\sin(10y))$.\\
\textit{Case B:} The model with parameters labeled \textit{Exact 2} is solved on the domain $\Omega = [-5,5]^2$ with $t\in[0,100]$. The initial conditions are chosen to be the perturbed unstable equilibrium as $\boldsymbol{c}(\boldsymbol{x},0) = (0.0527,0.0527)+\boldsymbol{\delta}$ where the perturbation, $\boldsymbol{\delta} = 10^{-4}\sin(||x||)/(0.001+||x||)(1,1)$. 

\begin{table}[]
\begin{tabular}{|c|ccc|cccccc|}
\hline
          & $\mathcal{D}_{12}$ & $\mathcal{D}_{23}$ & $\mathcal{D}_{13}$ & $R_{10}$ & $R_{11}$ & $R_{12}$ & $R_{20}$ & $R_{21}$ & $R_{22}$ \\ \hline
 Exact A & 5.00e0 & 5.00e0 & 5.00e-2 & 1.00e-1 & -2.00e+1 & 7.20e+3 & 1.00e0 & 0.00e0 &  -7.20e+3 \\
 Model A1 & 5.00e0 & 5.00e0 & 8.59e-2 & 1.35e-1 &
 -2.77e+1 & 9.99e+3 & 7.65e-1 & 9.69e0 &
 -9.99e+3\\ \hline
 Exact B & 8.33e-1 & 6.80e-1 & 1.68e-1 & 0.00e0 & -2.00e+1 & 7.20e+3 & 0.00e0 & 2.00e+1 &  -7.20e+3 \\
 Model B1 & 1.19e0 & 9.43e-1 & 2.33e-1 & 5.82e-4 &
 -2.78e+1 & 9.99e+3 & 3.34e-4 & 2.78e+1 &
 -9.99e+3\\
  Model B2 & 1.17e0 & 9.44e-1 & 2.33e-1 & 2.32e-4 &
 -2.78e+1 & 9.99e+3 & 1.30e-4 & 2.78e+1 &
 -9.99e+3\\
  Model B3 & 1.16e0 & 9.44e-1 & 2.33e-1 & 1.25e-4 &
 -2.78e+1 & 9.99e+3 & 5.33e-5 & 2.78e+1 &
 -9.99e+3\\
    Model B4 & 3.58e0 & 9.80e-1 & 3.26e-1 & 4.55e-2 &
 -2.86e+1 & 9.99e+3 & -4.01e-2 & 2.85e+1 &
 -9.99e+3\\\hline
\end{tabular}
\caption{Original and inferred parameters for MS Diffusion and Schnackenberg reaction system.}
\label{tab:schnack}
\end{table}

\subsubsection{Recovery of patterns and stability diagrams}

\begin{figure}
    \centering
    \includegraphics[width=0.95\linewidth]{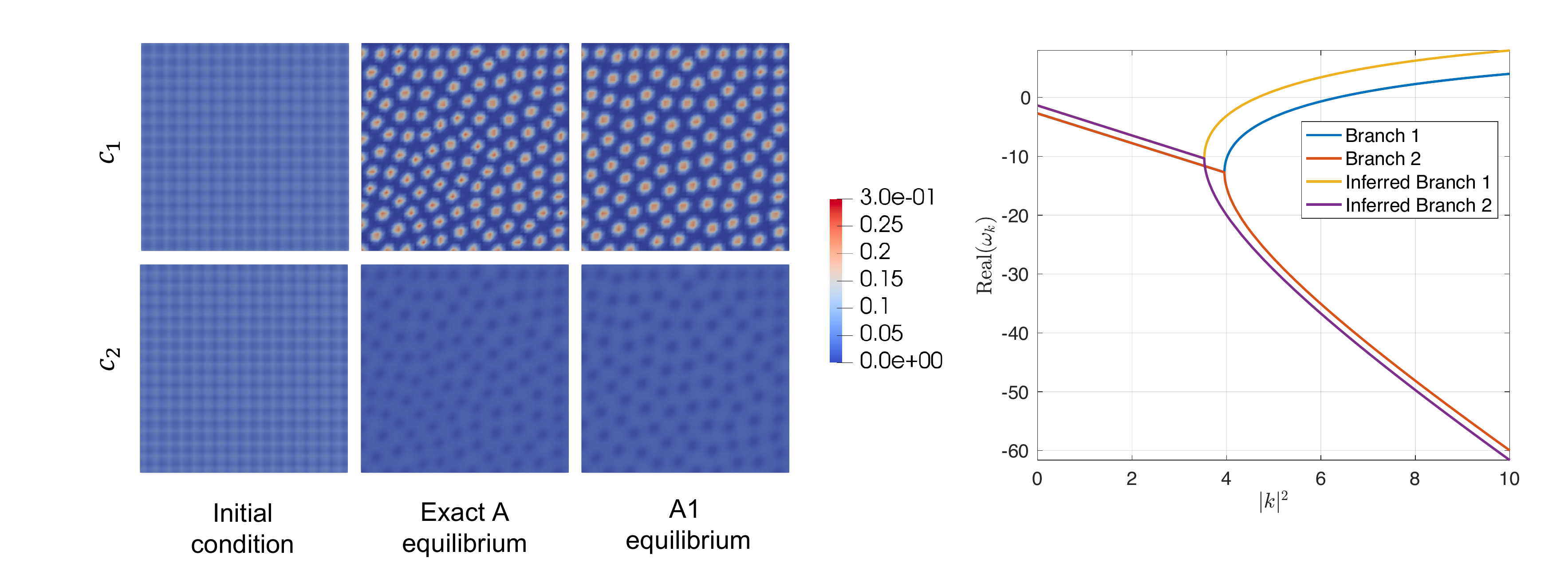}
    \caption{Comparison of original and inferred model for case 1. The left panel shows the state of the original and inferred model at $t=2$. The right panel shows the stability diagram for the original and inferred model.}
    \label{fig:case1_pattern}
\end{figure}

In this subsection, we will look at the results for inference of \textit{Case A}. 
We first observe the steady state behavior in this case (\cref{fig:case1_pattern}, left panel) forms \textit{spot}-patterns. The coefficients of the inferred model are provided in \cref{tab:schnack} with label, $A1$. We start with considering the reactive part of the inferred model and solve the reaction equations to estimate the homogeneous equilibrium solution, $\boldsymbol{c}(\boldsymbol{x}) = (0.05, 0.05)$. This solution is very close to the exact model's solution, $\boldsymbol{c}(\boldsymbol{x}) = (0.0527, 0.0527)$. Next, we solve the complete RD system with the inferred parameters A1 but with the initial conditions used to generate data. We observe that the equilibrium state of the inferred model (see \cref{fig:case1_pattern}) also shows a spotted pattern. The pattern formation process in the RD system is sensitive to initial conditions. Therefore, it is not possible to replicate the same steady-state state pattern from different models. However, we obtain similar patterns. We then construct and compare the stability diagram of the exact and the inferred models (see \cref{fig:case1_pattern}, right panel) and note that they are very similar. A similar stability diagram means that if two models start at the homogeneous unstable equilibrium, the same modes will grow and possibly lead to some steady state. However, the stability diagram doesn't directly determine the final patterns that are formed or the out-of-equilibrium evolution of the solution. 
This numerical study shows that the VSI technique can capture salient characteristics of the instabilities near the unstable equilibrium point and the long time patterning behavior of the system. There are two key takeaways from this numerical study:
\begin{enumerate}
    \item Similar patterns can be obtained from different models. This property will impact biological systems (discussed in the concluding remarks). 
    \item VSI can capture the characteristic features of instability and the long-term behavior or patterning in the RD system. However, VSI could not recover the coefficient of the original system. This phenomenon is attributed to the nonlinear optimization of \eqref{eq:VSI_loss}. The minimization problem has multiple local minima, and the algorithm identifies one of them as the solution. 
\end{enumerate}
In the following subsection, we will discuss ways to better inform the inference by choosing timestamps at which data is being collected.


\subsubsection{Effect of time slicing on inference}

\begin{figure}
    \centering
    \includegraphics[width=0.9\linewidth]{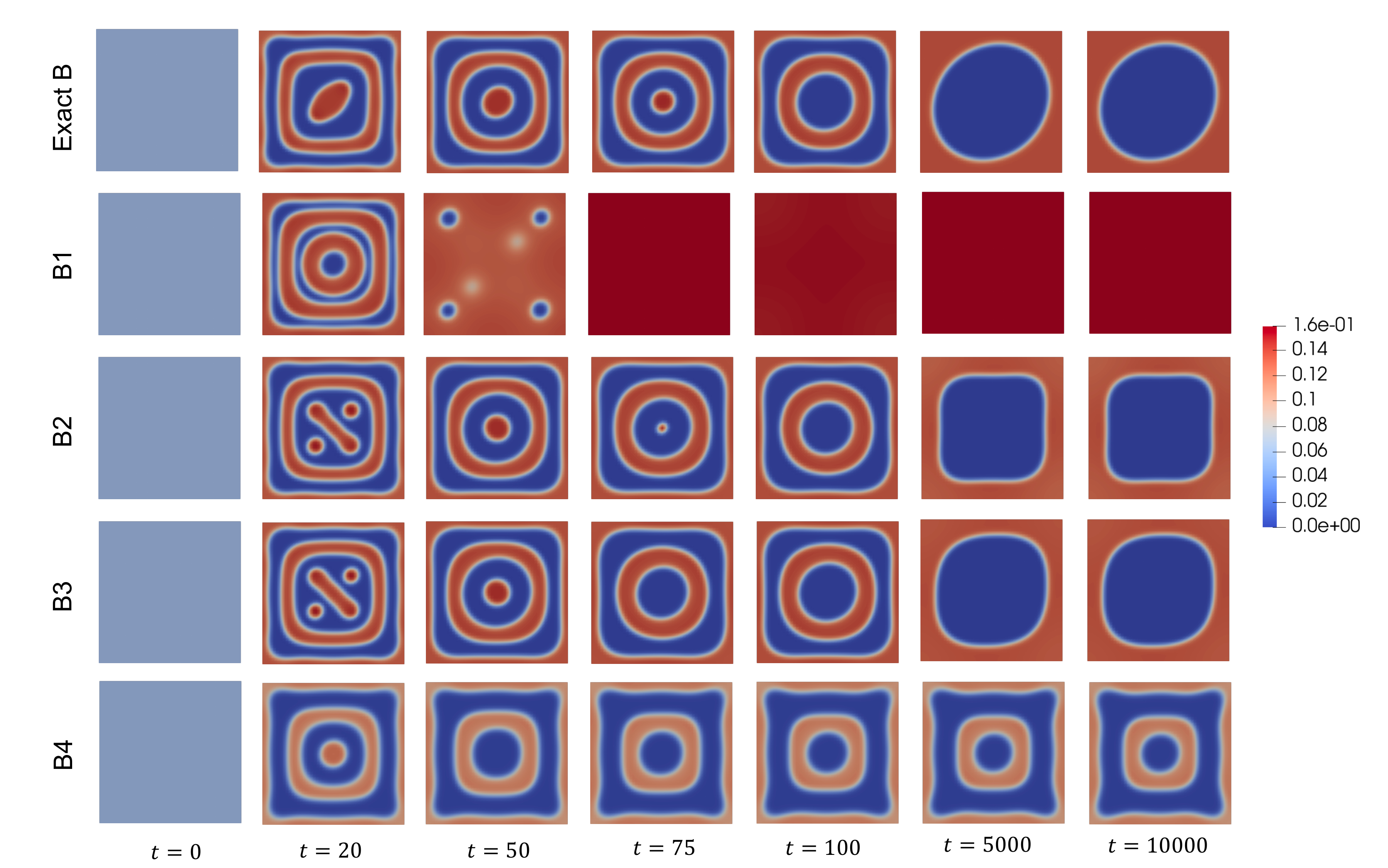}
    \caption{Spatiotemporal evolution of $c_1$ described by an RD system with parameters provided in \cref{tab:schnack}. }
    \label{fig:time_training}
\end{figure}

\begin{figure}
    \centering
    \includegraphics[width=0.6\linewidth]{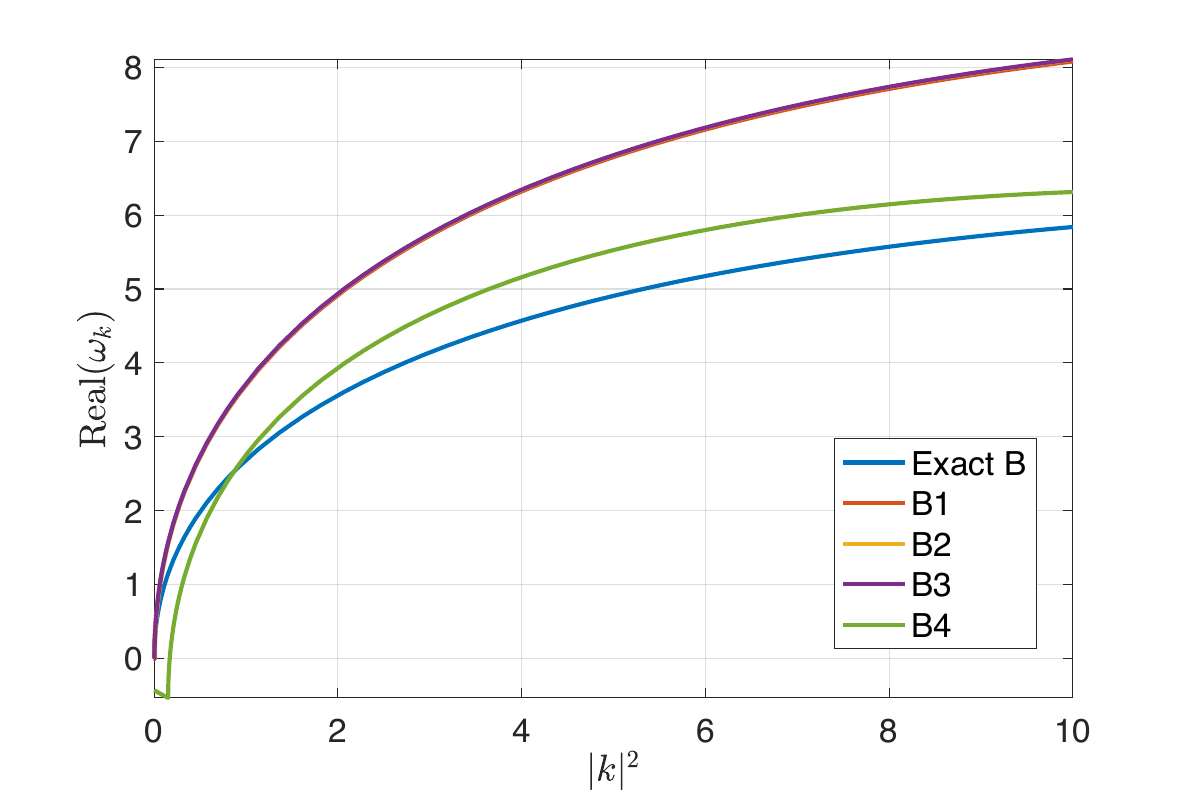}
    \caption{Unstable mode in the stability analysis for the models described in \cref{tab:schnack} (Exact B, B1, B2m B3, B4). All diagrams are evaluated at the equilibrium point $\boldsymbol{c}(\boldsymbol{x})=(0.0527, 0.0527)$. The curves corresponding to B1, B2 and B3 overlap and cannot be distinguished in this plot.} 
    \label{fig:time_training_stability}
\end{figure}

In this subsection, we will consider the inference of case B. We will solve this inference problem four times with data collected at different time points at $t\in \{0, \Delta t, 2\Delta t, \cdots n_T\Delta t\}$ where we consider the following cases: 
\begin{itemize}
    \item \textit{Sub-case B1}: simulation range: $t\in [0,20]$, time discretization $ n_T = 200, \Delta t = 0.1$
    \item \textit{Sub-case B2}: simulation range: $t\in [0,50]$, time discretization $ n_T = 500, \Delta t = 0.1$
    \item \textit{Sub-case B3}: simulation range: $t\in [0,100]$, time discretization $ n_T = 1000, \Delta t = 0.1$    
    \item \textit{Sub-case B4}: simulation range: $t\in [0,1]$, time discretization $ n_T = 100, \Delta t = 0.01$  
\end{itemize}
The numerical results for the inferred model are presented in \cref{tab:schnack}, labeled as B1-4. We again observe that in all cases, B1-4 inferred values differ from the original system. The solutions to the inferred models B1-4 for $t\in [0,10000]$ are provided in \cref{fig:time_training}. We reemphasize that cases B1, B2, B3 and B4 were trained with initial time series data up to $t=20$, $t=50$, $t=100$ and $t=1$ respectively. The forward solutions of the inferred models (presented in \cref{fig:time_training}) are carried out for a longer time range of $t\in[0,10000]$ to determine the steady state solution. We observe that all inferred models resemble the square patterns in the original data in the initial few time steps. However, sub-case B1 deviates from the data when it continues to evolve beyond its training range. Interestingly, the equilibrium solutions to sub-case B2 and sub-case B3 resemble the data, illustrating that these models can be extrapolated to longer times than the data with which they were trained. It is also seen that including a more extended range for data gathering B4 to B1 to B2 to B3 results in a better recovery of steady-state solution. However, when we compare the stability diagram of the inferred and exact models (see \cref{fig:time_training_stability}), we observe that the sub-cases B1, B2, and B3 are overpredicting the growth rates ($\text{Real}(\omega_k)$) for all wavenumbers $k$. Moreover, the stability diagrams for cases $B1$ to $B3$ are similar (overlapping in the \cref{fig:time_training_stability}). Meanwhile, the sub-case B4, which contains data for a small range of $t\in [0,1]$ with refined timesteps of $\Delta t = 0.01$, shows considerable improvement in recovering the stability diagram. There are two key takeaways from this study: 
\begin{enumerate}
    \item Increasing the time range allows VSI to identify model parameters that recover the patterns. 
    \item Refining time in the initial state close to the unstable equilibrium allows VSI to identify model parameters that recover the stability diagram
\end{enumerate}

\section{Concluding remarks}\label{sec:conclusion}

RD systems form a general class of PDEs that model pattern formation in biological systems. These models are typically studied in the context of Fick-based diffusion that defines a linear relationship between flux and the gradient of the concentration field. This idea is generalized for multicomponent transport by constructing flux as a linear combination of concentration gradients of each species. However, this relationship is only valid for low concentrations and, in general, is not thermodynamically consistent. MS diffusion provides a more general framework for multicomponent transport that is valid for arbitrary concentrations. We illustrated in \cref{sec:dunan_n_toor} that MS diffusion includes mechanisms for intermolecular force balance in the diffusion flux that allow us to describe phenomena like osmotic and reverse diffusion. Constructing an RD system with this mechanism will lead to a richer class of dynamics while being grounded in physics. We have also considered cases in \cref{sec:interpretation_pattern}, where at low concentrations, one can have slow and fast species that lead to pattern formation with the guiding physical principle of \textit{short-range activators coupled with long-range inhibitors}. We also showed that cross-diffusion can trigger these instabilities for pattern formation. MS-based diffusion allows a natural way to incorporate cross-diffusion at higher concentrations while maintaining thermodynamic consistency.

Meanwhile, simple reaction dynamics can capture a relatively broad class of patterns\cite{konda2010}. In our work, we observed spots and ring patterns with  Schnackenberg reaction dynamics. The MS-based RD system can be very quickly coupled with other reaction models studied in pattern formation, like the FitzHugh-Nagumo and the Gray-Scott models, offering a general framework for studying pattern formation while maintaining thermodynamic consistency. One should, however, practice caution when using these general reaction systems. The initial conditions and parameter space of this system must reflect the positivity-preserving property in the reaction dynamics required by the MS-based framework. The other restriction of this framework, the no-net production property, is automatically satisfied by the dynamics of the reduced system. At the same, this property offers more insight into the mechanics of pattern formation in these systems. 

\textit{No-net production property in MS-based RD system}: In a recent study by Braun et.al. \cite{brauns2020phase}, they illustrated Turing instabilities and patterning with mass-conserving reaction dynamics. 
\begin{align*}
    \frac{\partial c_1}{\partial t} = \Delta c_1 + f(c_1, c_2)\\
    \frac{\partial c_2}{\partial t} = \Delta c_2 - f(c_1, c_2) 
\end{align*}
the implication being that we get source-free transient diffusion for $c_1+c_2$. A central assumption in MS-based RD systems is that the total molar concentration remains constant when considering the complete $n-$component system.  As a consequence,  the ternary systems that we have illustrated in \cref{sec:inference_pattern} follow concentration-conserving reaction dynamics. Therefore, the patterns that we observe in this system are formed by redistribution and interchanging of species in the domain. This contrasts mechanisms studied in the general form of RD dynamics, where a local concentration change may alter the global concentration.

\textit{Implications in biological patterning and morphogenesis}: RD systems form one of the two primary schools of thought in the genesis of natural pattern formation, the other being mechanochemical systems\cite{garikipati2017perspectives,murray1988mechanochemical}. In the former approach, different species are treated as morphogens (or hypothetical chemicals) that can react and diffuse to form patterns. In many biological systems, these morphogens may interact via physical contact; for instance, the physics of pattern formation in tissues is dictated by the transport of a mixture of cells. Practical examples of such systems include embryogenesis, tissue differentiation, vasculogenesis, neural tube formation, invasive tumor growth\cite{garikipati2017perspectives,murray1988mechanochemical, gierer1972theory, czirok2013endothelial, invasive_tumor_growth}.
Modeling these phenomena must be done considering the high cell densities, warranting the need for more sophisticated modeling of diffusion mechanisms. As we discussed in \cref{sec:interpretation_pattern}, the spatial variations in densities may lead to different mechanisms for diffusion in different parts of the domain. 

MS-based RD system offers a general yet simple representation for modeling these complex biological behavior. The initial perturbation dependence of the final pattern also captures the uncertainty and variability we observe in nature. Moreover, the nonlinearity in the inference formulation shows a possible existence of multiple mechanisms (local minima in parameter space) that can explain pattern formation with these models. The class of patterns studied with this approach can be further broadened by including more pattern-forming reactive models like the FitzHugh-Nagumo model to determine the best reactive operators for the data. The VSI approach developed in the work can be integrated with operator selection techniques that we have presented in some previous works\cite{wang_system_2021,wang_variational_2019,wang2021variational}, allowing us to identify a parsimonious representation for reactive dynamics in a data-driven fashion.  In future, we will generalize this work to non-equimolar settings\cite{tai1980non} that will allow us to model a more general class of biological systems, such as the ones with significant cell proliferation and death.


\newpage
\pagestyle{bibliography}
\bibliography{main}

\begin{thebibliography}{10}
\expandafter\ifx\csname url\endcsname\relax
  \def\url#1{\texttt{#1}}\fi
\expandafter\ifx\csname urlprefix\endcsname\relax\def\urlprefix{URL }\fi
\providecommand{\bibinfo}[2]{#2}
\providecommand{\eprint}[2][]{\url{#2}}

\bibitem{garikipati2017perspectives}
\bibinfo{author}{Garikipati, K.}
\newblock \bibinfo{title}{Perspectives on the mathematics of biological patterning and morphogenesis}.
\newblock \emph{\bibinfo{journal}{Journal of the Mechanics and Physics of Solids}} \textbf{\bibinfo{volume}{99}}, \bibinfo{pages}{192--210} (\bibinfo{year}{2017}).

\bibitem{konda2010}
\bibinfo{author}{Kondo, S.} \& \bibinfo{author}{Miura, T.}
\newblock \bibinfo{title}{Reaction-diffusion model as a framework for understanding biological pattern formation}.
\newblock \emph{\bibinfo{journal}{Science}} \textbf{\bibinfo{volume}{329}}, \bibinfo{pages}{1616--1620} (\bibinfo{year}{2010}).
\newblock \urlprefix\url{https://www.science.org/doi/abs/10.1126/science.1179047}.
\newblock \eprint{https://www.science.org/doi/pdf/10.1126/science.1179047}.

\bibitem{turing1990chemical}
\bibinfo{author}{Turing, A.~M.}
\newblock \bibinfo{title}{The chemical basis of morphogenesis}.
\newblock \emph{\bibinfo{journal}{Bulletin of mathematical biology}} \textbf{\bibinfo{volume}{52}}, \bibinfo{pages}{153--197} (\bibinfo{year}{1990}).

\bibitem{green2015positional}
\bibinfo{author}{Green, J.~B.} \& \bibinfo{author}{Sharpe, J.}
\newblock \bibinfo{title}{Positional information and reaction-diffusion: two big ideas in developmental biology combine}.
\newblock \emph{\bibinfo{journal}{Development}} \textbf{\bibinfo{volume}{142}}, \bibinfo{pages}{1203--1211} (\bibinfo{year}{2015}).

\bibitem{madzvamuse2015cross}
\bibinfo{author}{Madzvamuse, A.}, \bibinfo{author}{Ndakwo, H.~S.} \& \bibinfo{author}{Barreira, R.}
\newblock \bibinfo{title}{Cross-diffusion-driven instability for reaction-diffusion systems: analysis and simulations}.
\newblock \emph{\bibinfo{journal}{Journal of mathematical biology}} \textbf{\bibinfo{volume}{70}}, \bibinfo{pages}{709--743} (\bibinfo{year}{2015}).

\bibitem{vanag2009cross}
\bibinfo{author}{Vanag, V.~K.} \& \bibinfo{author}{Epstein, I.~R.}
\newblock \bibinfo{title}{Cross-diffusion and pattern formation in reaction--diffusion systems}.
\newblock \emph{\bibinfo{journal}{Physical Chemistry Chemical Physics}} \textbf{\bibinfo{volume}{11}}, \bibinfo{pages}{897--912} (\bibinfo{year}{2009}).

\bibitem{Gorban2010QuasichemicalMO}
\bibinfo{author}{Gorban, A.~N.}, \bibinfo{author}{Sargsyan, H.} \& \bibinfo{author}{Wahab, H.~A.}
\newblock \bibinfo{title}{Quasichemical models of multicomponent nonlinear diffusion}.
\newblock \emph{\bibinfo{journal}{Mathematical Modelling of Natural Phenomena}} \textbf{\bibinfo{volume}{6}}, \bibinfo{pages}{184--262} (\bibinfo{year}{2010}).

\bibitem{KRISHNA1997861}
\bibinfo{author}{Krishna, R.} \& \bibinfo{author}{Wesselingh, J.}
\newblock \bibinfo{title}{The maxwell-stefan approach to mass transfer}.
\newblock \emph{\bibinfo{journal}{Chemical Engineering Science}} \textbf{\bibinfo{volume}{52}}, \bibinfo{pages}{861--911} (\bibinfo{year}{1997}).
\newblock \urlprefix\url{https://www.sciencedirect.com/science/article/pii/S0009250996004587}.

\bibitem{darken1948diffusion}
\bibinfo{author}{Darken, L.~S.}
\newblock \bibinfo{title}{Diffusion, mobility and their interrelation through free energy in binary metallic systems}.
\newblock \emph{\bibinfo{journal}{Trans. Aime}} \textbf{\bibinfo{volume}{175}}, \bibinfo{pages}{184--201} (\bibinfo{year}{1948}).

\bibitem{vignes1966diffusion}
\bibinfo{author}{Vignes, A.}
\newblock \bibinfo{title}{Diffusion in binary solutions. variation of diffusion coefficient with composition}.
\newblock \emph{\bibinfo{journal}{Industrial \& Engineering Chemistry Fundamentals}} \textbf{\bibinfo{volume}{5}}, \bibinfo{pages}{189--199} (\bibinfo{year}{1966}).

\bibitem{DAGOSTINO2011}
\bibinfo{author}{D'Agostino, C.}, \bibinfo{author}{Mantle, M.}, \bibinfo{author}{Gladden, L.} \& \bibinfo{author}{Moggridge, G.}
\newblock \bibinfo{title}{Prediction of binary diffusion coefficients in non-ideal mixtures from nmr data: Hexane–nitrobenzene near its consolute point}.
\newblock \emph{\bibinfo{journal}{Chemical Engineering Science}} \textbf{\bibinfo{volume}{66}}, \bibinfo{pages}{3898--3906} (\bibinfo{year}{2011}).
\newblock \urlprefix\url{https://www.sciencedirect.com/science/article/pii/S0009250911003150}.

\bibitem{duncan1962experimental}
\bibinfo{author}{Duncan, J.~B.} \& \bibinfo{author}{Toor, H.}
\newblock \bibinfo{title}{An experimental study of three component gas diffusion}.
\newblock \emph{\bibinfo{journal}{AIChE Journal}} \textbf{\bibinfo{volume}{8}}, \bibinfo{pages}{38--41} (\bibinfo{year}{1962}).

\bibitem{Bothe2011}
\bibinfo{author}{Bothe, D.}
\newblock \emph{\bibinfo{title}{On the Maxwell-Stefan Approach to Multicomponent Diffusion}}, \bibinfo{pages}{81--93} (\bibinfo{publisher}{Springer Basel}, \bibinfo{address}{Basel}, \bibinfo{year}{2011}).
\newblock \urlprefix\url{https://doi.org/10.1007/978-3-0348-0075-4_5}.

\bibitem{ramm2021diffusiophoretic}
\bibinfo{author}{Ramm, B.} \emph{et~al.}
\newblock \bibinfo{title}{A diffusiophoretic mechanism for atp-driven transport without motor proteins}.
\newblock \emph{\bibinfo{journal}{Nature Physics}} \textbf{\bibinfo{volume}{17}}, \bibinfo{pages}{850--858} (\bibinfo{year}{2021}).

\bibitem{yang2016dynamic}
\bibinfo{author}{Yang, K.}
\newblock \bibinfo{title}{Dynamic binary protein adsorption in ion-exchange media depicted with a parallel diffusion model derived from maxwell--stefan theory}.
\newblock \emph{\bibinfo{journal}{Chemical Engineering Science}} \textbf{\bibinfo{volume}{139}}, \bibinfo{pages}{163--172} (\bibinfo{year}{2016}).

\bibitem{sun2008analysis}
\bibinfo{author}{Sun, Y.} \& \bibinfo{author}{Yang, K.}
\newblock \bibinfo{title}{Analysis of mass transport models based on maxwell--stefan theory and fick's law for protein uptake to porous anion exchanger}.
\newblock \emph{\bibinfo{journal}{Separation and Purification Technology}} \textbf{\bibinfo{volume}{60}}, \bibinfo{pages}{180--189} (\bibinfo{year}{2008}).

\bibitem{membranes12100942}
\bibinfo{author}{Mazumder, A.}, \bibinfo{author}{Dobyns, B.~M.}, \bibinfo{author}{Howard, M.~P.} \& \bibinfo{author}{Beckingham, B.~S.}
\newblock \bibinfo{title}{Theoretical and experimental considerations for investigating multicomponent diffusion in hydrated, dense polymer membranes}.
\newblock \emph{\bibinfo{journal}{Membranes}} \textbf{\bibinfo{volume}{12}} (\bibinfo{year}{2022}).
\newblock \urlprefix\url{https://www.mdpi.com/2077-0375/12/10/942}.

\bibitem{fowler2018maxwell}
\bibinfo{author}{Fowler, K.}, \bibinfo{author}{Connolly, P.~J.}, \bibinfo{author}{Topping, D.~O.} \& \bibinfo{author}{O'Meara, S.}
\newblock \bibinfo{title}{Maxwell--stefan diffusion: a framework for predicting condensed phase diffusion and phase separation in atmospheric aerosol}.
\newblock \emph{\bibinfo{journal}{Atmospheric Chemistry and Physics}} \textbf{\bibinfo{volume}{18}}, \bibinfo{pages}{1629--1642} (\bibinfo{year}{2018}).

\bibitem{herberg2017reaction}
\bibinfo{author}{Herberg, M.}, \bibinfo{author}{Meyries, M.}, \bibinfo{author}{Pr{\"u}ss, J.} \& \bibinfo{author}{Wilke, M.}
\newblock \bibinfo{title}{Reaction--diffusion systems of maxwell--stefan type with reversible mass-action kinetics}.
\newblock \emph{\bibinfo{journal}{Nonlinear Analysis}} \textbf{\bibinfo{volume}{159}}, \bibinfo{pages}{264--284} (\bibinfo{year}{2017}).

\bibitem{giacomazzi2008review}
\bibinfo{author}{Giacomazzi, E.}, \bibinfo{author}{Picchia, F.} \& \bibinfo{author}{Arcidiacono, N.}
\newblock \bibinfo{title}{A review of chemical diffusion: Criticism and limits of simplified methods for diffusion coefficient calculation}.
\newblock \emph{\bibinfo{journal}{Combustion Theory and Modelling}} \textbf{\bibinfo{volume}{12}}, \bibinfo{pages}{135--158} (\bibinfo{year}{2008}).

\bibitem{sindy}
\bibinfo{author}{Brunton, S.~L.}, \bibinfo{author}{Proctor, J.~L.} \& \bibinfo{author}{Kutz, J.~N.}
\newblock \bibinfo{title}{Discovering governing equations from data by sparse identification of nonlinear dynamical systems}.
\newblock \emph{\bibinfo{journal}{Proceedings of the National Academy of Sciences}} \textbf{\bibinfo{volume}{113}}, \bibinfo{pages}{3932--3937} (\bibinfo{year}{2016}).
\newblock \urlprefix\url{https://www.pnas.org/doi/abs/10.1073/pnas.1517384113}.
\newblock \eprint{https://www.pnas.org/doi/pdf/10.1073/pnas.1517384113}.

\bibitem{wang2021variational}
\bibinfo{author}{Wang, Z.}, \bibinfo{author}{Huan, X.} \& \bibinfo{author}{Garikipati, K.}
\newblock \bibinfo{title}{Variational system identification of the partial differential equations governing microstructure evolution in materials: Inference over sparse and spatially unrelated data}.
\newblock \emph{\bibinfo{journal}{Computer Methods in Applied Mechanics and Engineering}} \textbf{\bibinfo{volume}{377}}, \bibinfo{pages}{113706} (\bibinfo{year}{2021}).

\bibitem{nikolov2022ogden}
\bibinfo{author}{Nikolov, D.~P.} \emph{et~al.}
\newblock \bibinfo{title}{Ogden material calibration via magnetic resonance cartography, parameter sensitivity and variational system identification}.
\newblock \emph{\bibinfo{journal}{Philosophical Transactions of the Royal Society A}} \textbf{\bibinfo{volume}{380}}, \bibinfo{pages}{20210324} (\bibinfo{year}{2022}).

\bibitem{wang_system_2021}
\bibinfo{author}{Wang, Z.}, \bibinfo{author}{Carrasco-Teja, M.}, \bibinfo{author}{Zhang, X.}, \bibinfo{author}{Teichert, G.~H.} \& \bibinfo{author}{Garikipati, K.}
\newblock \bibinfo{title}{System {Inference} {Via} {Field} {Inversion} for the {Spatio}-{Temporal} {Progression} of {Infectious} {Diseases}: {Studies} of {COVID}-19 in {Michigan} and {Mexico}}.
\newblock \emph{\bibinfo{journal}{Archives of Computational Methods in Engineering}} \textbf{\bibinfo{volume}{28}}, \bibinfo{pages}{4283--4295} (\bibinfo{year}{2021}).
\newblock \urlprefix\url{https://link.springer.com/10.1007/s11831-021-09643-1}.

\bibitem{wang_variational_2019}
\bibinfo{author}{Wang, Z.}, \bibinfo{author}{Huan, X.} \& \bibinfo{author}{Garikipati, K.}
\newblock \bibinfo{title}{Variational system identification of the partial differential equations governing the physics of pattern-formation: {Inference} under varying fidelity and noise}.
\newblock \emph{\bibinfo{journal}{Computer Methods in Applied Mechanics and Engineering}} \textbf{\bibinfo{volume}{356}}, \bibinfo{pages}{44--74} (\bibinfo{year}{2019}).
\newblock \urlprefix\url{https://linkinghub.elsevier.com/retrieve/pii/S0045782519304037}.

\bibitem{BOTHE2023103818}
\bibinfo{author}{Bothe, D.} \& \bibinfo{author}{Étienne Druet, P.}
\newblock \bibinfo{title}{On the structure of continuum thermodynamical diffusion fluxes—a novel closure scheme and its relation to the maxwell–stefan and the fick–onsager approach}.
\newblock \emph{\bibinfo{journal}{International Journal of Engineering Science}} \textbf{\bibinfo{volume}{184}}, \bibinfo{pages}{103818} (\bibinfo{year}{2023}).
\newblock \urlprefix\url{https://www.sciencedirect.com/science/article/pii/S0020722523000095}.

\bibitem{bothe2015continuum}
\bibinfo{author}{Bothe, D.} \& \bibinfo{author}{Dreyer, W.}
\newblock \bibinfo{title}{Continuum thermodynamics of chemically reacting fluid mixtures}.
\newblock \emph{\bibinfo{journal}{Acta Mechanica}} \textbf{\bibinfo{volume}{226}}, \bibinfo{pages}{1757--1805} (\bibinfo{year}{2015}).

\bibitem{boudin2012mathematical}
\bibinfo{author}{Boudin, L.}, \bibinfo{author}{Grec, B.} \& \bibinfo{author}{Salvarani, F.}
\newblock \bibinfo{title}{A mathematical and numerical analysis of the maxwell-stefan diffusion equations}.
\newblock \emph{\bibinfo{journal}{Discrete and Continuous Dynamical Systems-Series B}} \textbf{\bibinfo{volume}{17}}, \bibinfo{pages}{1427--1440} (\bibinfo{year}{2012}).

\bibitem{dolfin}
\bibinfo{author}{Logg, A.} \& \bibinfo{author}{Wells, G.~N.}
\newblock \bibinfo{title}{Dolfin: Automated finite element computing}.
\newblock \emph{\bibinfo{journal}{ACM Trans. Math. Softw.}} \textbf{\bibinfo{volume}{37}} (\bibinfo{year}{2010}).
\newblock \urlprefix\url{https://doi.org/10.1145/1731022.1731030}.

\bibitem{logg2012automated}
\bibinfo{author}{Logg, A.}, \bibinfo{author}{Mardal, K.-A.} \& \bibinfo{author}{Wells, G.}
\newblock \emph{\bibinfo{title}{Automated solution of differential equations by the finite element method: The FEniCS book}}, vol.~\bibinfo{volume}{84} (\bibinfo{publisher}{Springer Science \& Business Media}, \bibinfo{year}{2012}).

\bibitem{scikit-learn}
\bibinfo{author}{Pedregosa, F.} \emph{et~al.}
\newblock \bibinfo{title}{Scikit-learn: Machine learning in {P}ython}.
\newblock \emph{\bibinfo{journal}{Journal of Machine Learning Research}} \textbf{\bibinfo{volume}{12}}, \bibinfo{pages}{2825--2830} (\bibinfo{year}{2011}).

\bibitem{KRISHNA1997}
\bibinfo{author}{Krishna, R.} \& \bibinfo{author}{Wesselingh, J.}
\newblock \bibinfo{title}{The maxwell-stefan approach to mass transfer}.
\newblock \emph{\bibinfo{journal}{Chemical Engineering Science}} \textbf{\bibinfo{volume}{52}}, \bibinfo{pages}{861--911} (\bibinfo{year}{1997}).
\newblock \urlprefix\url{https://www.sciencedirect.com/science/article/pii/S0009250996004587}.

\bibitem{brauns2020phase}
\bibinfo{author}{Brauns, F.}, \bibinfo{author}{Halatek, J.} \& \bibinfo{author}{Frey, E.}
\newblock \bibinfo{title}{Phase-space geometry of mass-conserving reaction-diffusion dynamics}.
\newblock \emph{\bibinfo{journal}{Physical Review X}} \textbf{\bibinfo{volume}{10}}, \bibinfo{pages}{041036} (\bibinfo{year}{2020}).

\bibitem{murray1988mechanochemical}
\bibinfo{author}{Murray, J.~D.}, \bibinfo{author}{Maini, P.~K.} \& \bibinfo{author}{Tranquillo, R.~T.}
\newblock \bibinfo{title}{Mechanochemical models for generating biological pattern and form in development}.
\newblock \emph{\bibinfo{journal}{Physics Reports}} \textbf{\bibinfo{volume}{171}}, \bibinfo{pages}{59--84} (\bibinfo{year}{1988}).

\bibitem{gierer1972theory}
\bibinfo{author}{Gierer, A.} \& \bibinfo{author}{Meinhardt, H.}
\newblock \bibinfo{title}{A theory of biological pattern formation}.
\newblock \emph{\bibinfo{journal}{Kybernetik}} \textbf{\bibinfo{volume}{12}}, \bibinfo{pages}{30--39} (\bibinfo{year}{1972}).

\bibitem{czirok2013endothelial}
\bibinfo{author}{Czirok, A.}
\newblock \bibinfo{title}{Endothelial cell motility, coordination and pattern formation during vasculogenesis}.
\newblock \emph{\bibinfo{journal}{Wiley Interdisciplinary Reviews: Systems Biology and Medicine}} \textbf{\bibinfo{volume}{5}}, \bibinfo{pages}{587--602} (\bibinfo{year}{2013}).

\bibitem{invasive_tumor_growth}
\bibinfo{author}{Khain, E.} \& \bibinfo{author}{Sander, L.~M.}
\newblock \bibinfo{title}{Dynamics and pattern formation in invasive tumor growth}.
\newblock \emph{\bibinfo{journal}{Phys. Rev. Lett.}} \textbf{\bibinfo{volume}{96}}, \bibinfo{pages}{188103} (\bibinfo{year}{2006}).
\newblock \urlprefix\url{https://link.aps.org/doi/10.1103/PhysRevLett.96.188103}.

\bibitem{tai1980non}
\bibinfo{author}{Tai, R.~C.}, \bibinfo{author}{Chang, H.-K.} \& \bibinfo{author}{Farhi, L.~E.}
\newblock \bibinfo{title}{Non-equimolar counter-diffusion in ternary gas systems}.
\newblock \emph{\bibinfo{journal}{Respiration Physiology}} \textbf{\bibinfo{volume}{40}}, \bibinfo{pages}{253--267} (\bibinfo{year}{1980}).

\end{thebibliography}

\end{document}